\documentclass[table]{ccjnl}
\usepackage{lipsum,amsmath}
\usepackage{cuted}
\usepackage{bm}
\usepackage{epstopdf}
\usepackage{graphicx,psfrag,color}
\usepackage{subfigure}
\usepackage{tabularx}
\graphicspath{{figures/}}
\DeclareMathOperator*{\argmax}{arg\,max}
\title{Radar Sensing via OTFS Signaling}
\author{Kecheng Zhang\inst{1}, Zhongjie Li\inst{1}, Weijie Yuan\inst{1,*}, Yunlong Cai\inst{2}, Feifei Gao\inst{3},\corinfo{yuanwj@sustech.edu.cn}}
\receiveddate{xxx}
\reviseddate{xxx}
\Editor{}

\address[1]{Department of Electronic and Electrical Engineering, Southern University of Science and Technology, Shenzhen 518055, China}
\address[2]{College of Information Science and Electronic Engineering, Zhejiang University, Hangzhou 310027, China}
\address[3]{Department of Automation, Tsinghua University, Beijing 100084, China}

\begin{document}

\maketitle

\begin{abstract}
    By multiplexing information symbols in the delay-Doppler (DD) domain, orthogonal time frequency space (OTFS) is a promising candidate for future wireless communication in high-mobility scenarios. In addition to the superior communication performance, OTFS is also a natural choice for radar sensing since the primary parameters (range and velocity of targets) in radar signal processing can be inferred directly from the delay and Doppler shifts. Though there are several works on OTFS radar sensing, most of them consider the integer parameter estimation only, while the delay and Doppler shifts are usually fractional in the real world. In this paper, we propose a two-step method to estimate the fractional delay and Doppler shifts. We first perform the two-dimensional (2D) correlation between the received and transmitted DD domain symbols to obtain the integer parts of the parameters. Then a difference-based method is implemented to estimate the fractional parts of delay and Doppler indices. Meanwhile, we implement a target detection method based on a generalized likelihood ratio test since the number of potential targets in the sensing scenario is usually unknown. The simulation results show that the proposed method can obtain the delay and Doppler shifts accurately and get the number of sensing targets with a high detection probability.
    \keywords{OTFS; Radar sensing; delay-Doppler (DD) domain; Fractional delay and Doppler}
    \keywords{OTFS; 6G; delay-Doppler (DD) domain}
\end{abstract}
\section{Introduction}
\label{Introduction}
The future wireless systems, such as the sixth-generation (6G) wireless networks, are expected to provide both communication and sensing services simultaneously\cite{ISAC1}. It is believed that sensing will play an important role together with communications in the next-generation wireless network \cite{ISAC2}. For example, the future autonomous vehicle will receive a lot of information, including maps in high resolution and real-time navigating information, from the base station (BS). Meanwhile, the vehicle itself should have robust and high-accuracy sensing ability to avoid obstacles \cite{ISAC1}, which implies the implementation of integrated sensing and communication (ISAC).

To realize ISAC, several studies consider orthogonal frequency division multiplexing (OFDM) as the transmitted signal waveform \cite{OFDM1, OFDM2}. OFDM has good robustness and high spectral efficiency under time-invariant frequency selective channels \cite{OFDM1}, and low detection complexity for radar target sensing. However, OFDM suffers from several drawbacks such as Doppler intolerance and high peak-to-average power ratio (PAPR). In high-mobility environments, the communication performance of OFDM will be degraded significantly, since the channels in high-mobility environments are fast time-varying, and the orthogonality between OFDM subcarriers will be broken due to severe Doppler effect. Moreover, it will also be challenging to implement OFDM in high-frequency communication systems due to the high PAPR \cite{OFDM3}. Under these conditions, it is necessary to develop a new waveform that can support both robust radar sensing and high-quality wireless communications in high-mobility environments.

A new modulation scheme called orthogonal time frequency space (OTFS) modulation was proposed recently for reliable communications over multi-path channels in which the delay and Doppler shifts differed in each path. By describing the channel in the delay-Doppler (DD) domain \cite{OTFS0, OTFS_OFDM_compare}, OTFS provided an alternative representation of the fast time-varying channel produced by moving objects in wireless environments. With this representation, the information symbols were modulated over the two-dimensional DD domain, which has extra benefits compared to the traditional time-frequency (TF) domain modulation scheme. For example, compared to the fast time-varying channel in the TF domain, the channel in the DD domain is usually sparse and quasi-static \cite{OTFS1, Low_PAPR} in high-mobility wireless environments, which can simplify the channel estimation procedure and provide robust communication performance \cite{OTFS2}. Moreover, the range and velocity parameters of targets can be directly infferred from delay and Doppler shifts, which are essential parameters in radar signal processing, making OTFS a natural choice for realizing ISAC. In our previous work \cite{zhang2023radar}, we have unveiled that the OTFS demodulation is exactly the range-Doppler matrix calculation procedure in radar sensing. The complexity of the ISAC system will be reduced greatly since the sensing and communication can be done through the same calculation process if the OTFS signaling is implemented.

There have been many researches towards the communication via OTFS waveform \cite{OTFS4, SimpleBayesian, WORK1}, which showed the superior communication performance of OTFS under high-mobility scenario, but there are much less studies \cite{WORK1, WORK2, WORK3} about OTFS radar sensing. For example, in \cite{WORK1}, the authors proposed an OTFS-based ISAC system using the maximum likelihood (ML) algorithm to estimate the range and velocity of targets. The proposed system can maintain a superior communication performance compared to OFDM and achieve the radar estimation performance bound at the same time. The work \cite{WORK2} proposed an OTFS-based radar system using the matched-filter algorithm for target sensing. By placing randomly generated data symbols with the same power in the whole OTFS frame, the potential targets together with the associated ranges and velocities can be estimated via a matched-filter algorithm. In \cite{WORK3}, a generalized likelihood ratio test-based detector using OTFS waveform was proposed. Different from the two works mentioned above, this work performs sensing via the time-domain OTFS signal directly. The results show that the estimation performance is improved compared to the standard fast Fourier transform (FFT) based OFDM radar.

However, the potential of OTFS radar sensing requires further exploration. To achieve good performance, the calculation complexity of the algorithm in \cite{WORK1} is cubic increased with the frame size, which is relatively high. In the proposed algorithm in \cite{WORK2}, both the delay and Doppler shifts are considered as integers. However, this may not be practical in real-world wireless networks where the duration of one OTFS frame is limited, which can result in insufficient Doppler resolution. Hence, the fractional Doppler effect is inevitable in reality. In order to achieve accurate radar sensing, it is necessary to study the use of OTFS in the presence of off-grid targets, although integer delay is typically assumed sufficient for communication design. Therefore, it is necessary to take into account both fractional delay and Doppler shifts when using OTFS for radar sensing. The need to address these issues motivates the development of a method for radar sensing using OTFS in the presence of fractional delay and Doppler shifts.

In this paper, a two-step method for the fractional delay and Doppler indices estimation is proposed with the inspiration of the pulse compression technique in radar sensing \cite{RADAR_AND_OTFS0}. In the ISAC scenario, the transmitted OTFS frame is designed to provide both communication and sensing services, which means the delay and Doppler domain are fully occupied by information symbols. Under this condition, though the received data is demodulated into the DD domain, it is hard to locate the delay and Doppler shifts of targets due to the overlapped responses from other data symbol. Thus, we perform a 2D correlation-based method to first obtain the integer parts of the delay and Doppler parameters. Then, we propose a difference-based method to estimate the fractional parts of the parameters. Since the number of potential targets is usually unknown in the sensing scenario, we propose a target detection algorithm based on the generalized likelihood ratio test (GLRT). The simulation results show that the proposed algorithm can estimate the delay and Doppler shifts associated with multiple targets accurately.

\textit{Notations}: The boldface lowercase letter and boldface capital letter denote the vector and the matrix, respectively. The superscript T, *, H represent for transpose, conjugate, and the Hermittan operations, respectively; $|\cdot|$ denotes the modulus of a complex number or the cardinality of a set; $\parallel\cdot\parallel$ denotes the $L^2$ norm; $\mathbb{E}[x]$ denotes the expectation of $x$; $\partial$ is the partial derivative operator; the notation $\mathcal{O}$ is an asymptotic notation describing the order of errorness; $[\cdot]_{N}$ denotes the modulo operation with divisor $N$.
\section{System Model}
\label{OTFS}
In this section, we first recap the basic concepts in OTFS and then present the input-output relation of OTFS symbols in the delay-Doppler (DD) domain.
\subsection{Basic Concepts of OTFS}
Let $N\in\mathbb{N}$ and $M\in\mathbb{N}$ denote the number of time slots and subcarriers of an OTFS frame, respectively, where $\mathbb{N}$ is the positive integer set. The discrete delay-Doppler plane is denoted as
\begin{equation}
    \resizebox{0.9\hsize}{!}{$
            \mathbf{\Gamma}=\left\{\left(\frac{k}{NT},\frac{l}{M\Delta f}|k\in[0,N-1],l\in[0,M-1]\right)\right\}$}
\end{equation}

Denote the duration of one timeslot as $T$, and the subcarrier spacing as $\Delta f$, respectively, resulting in the duration of one OTFS frame of $NT$ and the occupied bandwidth of the whole system of $M\Delta f$. The resolutions towards the delay and Doppler shifts are $\frac{1}{M\Delta f}$ and $\frac{1}{NT}$, respectively.

The cross-ambiguity function for the pulse shaping filter at the transmitter $g_{tx}(t)$ and the receiver $g_{rx}$ can be written as
\begin{equation}
    A_{g_{rx}g_{tx}}\left(t,f\right)\overset{\Delta}{=}\int_{t}g_{tx}(t^{\prime})g_{rx}^{*}(t^{\prime}-t)e^{j2\pi\nu(t^{\prime}-t)}dt^{\prime}.
\end{equation}
For ease of exposition, we focus on the case where the transmitter and receiver of OTFS are applied by ideal pulse shaping filters \cite{OTFS_Window_Design}, which have the bi-orthogonal property \cite{OTFS0} as shown below
\begin{equation}
    \begin{aligned}
        A & _{g_{rx}g_{tx}}(t,f)|_{t=(n-n^{\prime})T-\frac{l_{\tau}}{M\Delta f},(m-m^{\prime})\Delta f-\frac{k_{\nu}}{NT}} \\
          & =\delta[n]\delta[m]q_{\tau_{max}}(t-nT)q_{\nu_{max}}(f-m\Delta f)
    \end{aligned}
\end{equation}
where $q_{a}(t)=1$ for $x\in(-a,a)$ and zero otherwise.
\subsection{Input-output Relation of OTFS Transmission}
\begin{figure}\centering
    \includegraphics[width=0.95\columnwidth]{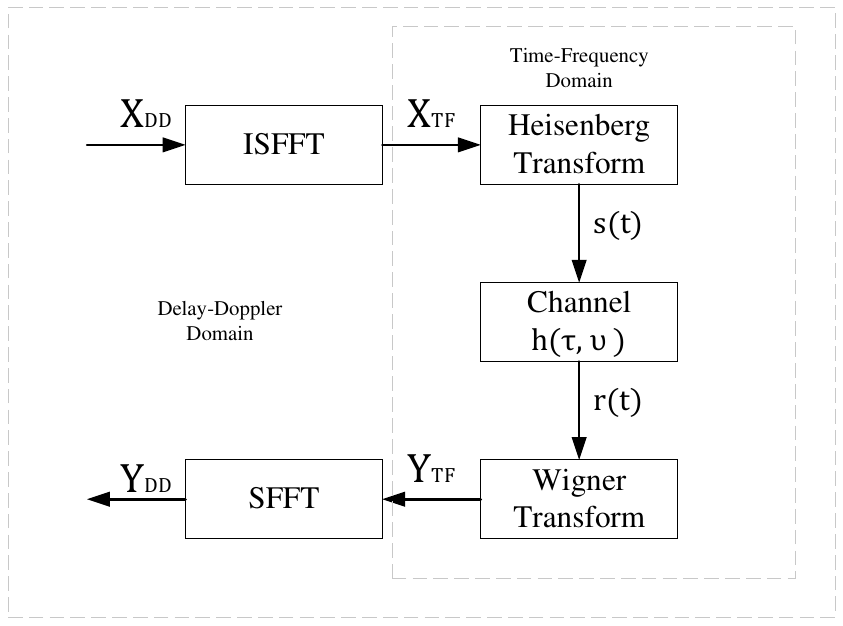}
    \caption{The input-output relation of OTFS transmission.}
    \label{io_relation}
\end{figure}
The OTFS input-output relation is summarized in Figure \ref{io_relation}. Below are the detailed description of the OTFS transmission procedure. The information bits sequence is mapped into a symbol set in the DD domain, $\left\{X_{\text{DD}}[k,l]|k=0,..., N-1,l=0,..., M-1\right\}$, where $k$ and $l$ are the Doppler and delay indicies respectively. $X_{\text{DD}}\in\mathbb{A}$, where $\mathbb{A}=\left\{a_{i}\in\mathbb{C}|i=1,2,...,|A|\right\}$ is the set of modulation alphabet (e.g. QAM). The OTFS symbols are transformed from the DD domain into the time-frequency (TF) domain via the inverse symplectic finite Fourier transform (ISFFT), which is described below
\begin{equation}
    X_{\text{TF}}[n,m]=\sum_{k=0}^{N-1}\sum_{l=0}^{M-1}X_{\text{DD}}[k,l]e^{j2\pi\left(\frac{nk}{N}-\frac{ml}{M}\right)},
\end{equation}
for $n=0,...,N-1$ and $m=0,...,M-1$.

Then the TF domain modulator converts the symbols $X_{\text{TF}}[n,m]$ to a continuous time waveform $s(t)$ with the transmitter pulse shaping filter $g_{tx}(t)$ via the Heisenberg transform \cite{OTFS0}
\begin{equation}
    s(t)=\sum_{n=0}^{N-1}\sum_{m=0}^{M-1}X_{\text{TF}}[n,m]g(t-nT)e^{j2\pi m\Delta f(t-nT)}.
\end{equation}

After passing a time-varying channel, the signal $s(t)$ arrives at the receiver. The received signal $r(t)$ is given by
\begin{equation}
    r(t)=\int\int h(\tau,\nu)s(t-\tau)e^{j2\pi\nu(t-\tau)}d\tau d\nu + w(t),
\end{equation}
where $w(t)$ is the additive white Gaussian noise (AWGN) process with one sider power spectral density (PSD) $\mathcal{N}_{0}$, $h(\tau,\nu)\in\mathbb{C}$ is the complex baseband channel impulse response, in which $\tau$ and $\nu$ represents the channel response to an impulse with delay $\tau$ and Doppler shift $\nu$, given by
\begin{equation}
    h(\tau,\nu)=\sum_{i=1}^{P}h_{i}\delta(\tau-\tau_{i})\delta(\nu-\nu_{i}),
\end{equation}
where $P\in\mathbb{Z}$ denotes the targets number in the sensing field, and $h_{i}$, $\tau_{i}$ and $\nu_{i}$ denote the reflection coefficient, delay, and Doppler shift associated with the $i$th target, respectively. Denote the range and relative velocity of the $i$th target as $R_{i}$ and $v_{i}$, then the round-trip delay $\tau_{i}\in\mathbb{R}$ and the Doppler frequency $\nu_{i}\in\mathbb{R}$ are represented as
\begin{equation}
    \label{nu_tau_def}
    \tau_{i}=\frac{2R_{i}}{c}=\frac{l_{\tau_{i}}}{M\Delta f},\nu_{i}=\frac{2f_{c}v_{i}}{c}=\frac{k_{\nu_{i}}}{NT},
\end{equation}
where $c$ is the speed of light and $f_{c}$ is the carrier frequency, $l_{\tau_{i}}\in\mathbb{R}$ and $k_{\nu_{i}}\in\mathbb{R}$ are the delay and Doppler indices of $i$th target. In this paper, we consider fractional delay and Doppler indices, and represent the delay and Doppler indices as $l_{\tau_{i}}=l_{i}+\iota_{i}$ and $k_{\nu_{i}}=k_{i}+\kappa_{i}$, where $l_{i}\in\mathbb{Z}$ and $k_{i}\in\mathbb{Z}$ are the integer parts of the indices associated with the $i$th target, and $\iota_{i}\in\mathbb{R}$ and $\kappa_{i}\in\mathbb{R}$ are the fractional part.

At the receiver, the received signal $r(t)$ is transformed into the TF domain by Wigner transform \cite{OTFS0}, which can also be expressed as the cross-ambiguity function $A_{g_{rx},r}(t,f)$ between the receiver pulse shaping filter and the received signal, given by
\begin{equation}
    \begin{aligned}
         & Y_{\text{TF}}(t,f)=A_{g_{rx},r}(t,f)                                                                \\
         & \overset{\Delta}{=}\int g_{rx}^{*}(t^{\prime}-t)r(t^{\prime})e^{-j2\pi f(t^{\prime}-t)}dt^{\prime}.
    \end{aligned}
\end{equation}

By sampling $Y(t,f)$, we have the discrete output as
\begin{equation}
    Y_{\text{TF}}[n,m]=Y_{\text{TF}}(t,f)|_{t=nT,f=m\Delta f},
\end{equation}
for $n=0,...,N-1$ and $m=0,...M-1$. Then, by applying SFFT on $Y_{\text{TF}}[n,m]$, we have the symbols $Y_{\text{DD}}[k,l]$ in the delay-Doppler domain
\begin{equation}
    Y_{\text{DD}}[k,l]=\frac{1}{\sqrt[]{NM}}\sum_{n=0}^{N-1}\sum_{m=0}^{M-1}Y_{\text{TF}}[n,m]e^{-j2\pi\left(\frac{nk}{N}-\frac{ml}{M}\right)}.
\end{equation}

Under the case of ideal pulse shaping filter \cite{raviteja2018interference}, the input-output relationship of OTFS symbols is given by
\begin{equation}
    \label{output_DD}
    \begin{aligned}
        Y_{\text{DD}}[k,l]=\sum_{n=0}^{N-1}\sum_{m=0}^{M-1} & X_{\text{DD}}[n,m]h_{\omega}[k-n,l-m] \\
                                                            & +Z_{\text{DD}}[k,l],
    \end{aligned}
\end{equation}
where $Z_{\text{DD}}\sim\mathcal{CN}(0,\sigma^{2}\mathbf{\textit{I}})$ is the effective noise in the DD domain, and $h_{\omega}[k,l]$ is the effective channel in DD domain obtained by
\begin{equation}
    \label{effective_channel}
    h_{\omega}[k,l]=\sum_{i=1}^{P}h_{i}\omega(k-k_{\nu_{i}},l-l_{\tau_{i}})e^{-j2\pi\nu_{i}\tau_{i}},
\end{equation}
where $\omega(k-k_{\nu_{i}},l-l_{\tau_{i}})$ is the sampling function, and $\nu_{i}$ and $\tau_{i}$ are given in (\ref{nu_tau_def}). By adopting the rectangular window \cite{OTFS_Window_Design} at both the transmitter and receiver in the TF domain, the sampling function is simplified to be
\begin{equation}
    \label{sampling_function}
    \begin{aligned}
         & \omega(k-k_{\nu_{i}},l-l_{\tau_{i}})=\mathcal{G}(k-k_{\nu_{i}})\mathcal{F}(l-l_{\tau_{i}}),                                      \\
         & \mathcal{G}(k-k_{\nu_{i}})\overset{\Delta}{=}\frac{1}{N}\sum_{k^{\prime}=0}^{N-1}e^{-j2\pi\frac{(k-k_{\nu_{i}})k^{\prime}}{N}},  \\
         & \mathcal{F}(l-l_{\tau_{i}})\overset{\Delta}{=}\frac{1}{M}\sum_{l^{\prime}=0}^{M-1}e^{j2\pi\frac{(l-l_{\tau_{i}})l^{\prime}}{M}}.
    \end{aligned}
\end{equation}

According to \cite{raviteja2018interference}, the non-zero entries of the DD domain effective channel are localized identically for both the ideal and rectangular pulse shaping filter, and they only differ from an additional phase offset. Therefore, the designed channel estimation algorithm in this paper can be easily extended to the case of a rectangular pulse shaping filter.
\section{OTFS Based Radar Sensing}\label{PROPOSED}
\vspace{-0.1em}
In this section, we introduce the proposed target detection method and the 2D correlation-based method for parameter estimation.
\begin{figure} \centering
    \subfigure[Before performing the 2D correlation.] {\label{received_matrix}
        \includegraphics[width=0.46\columnwidth]{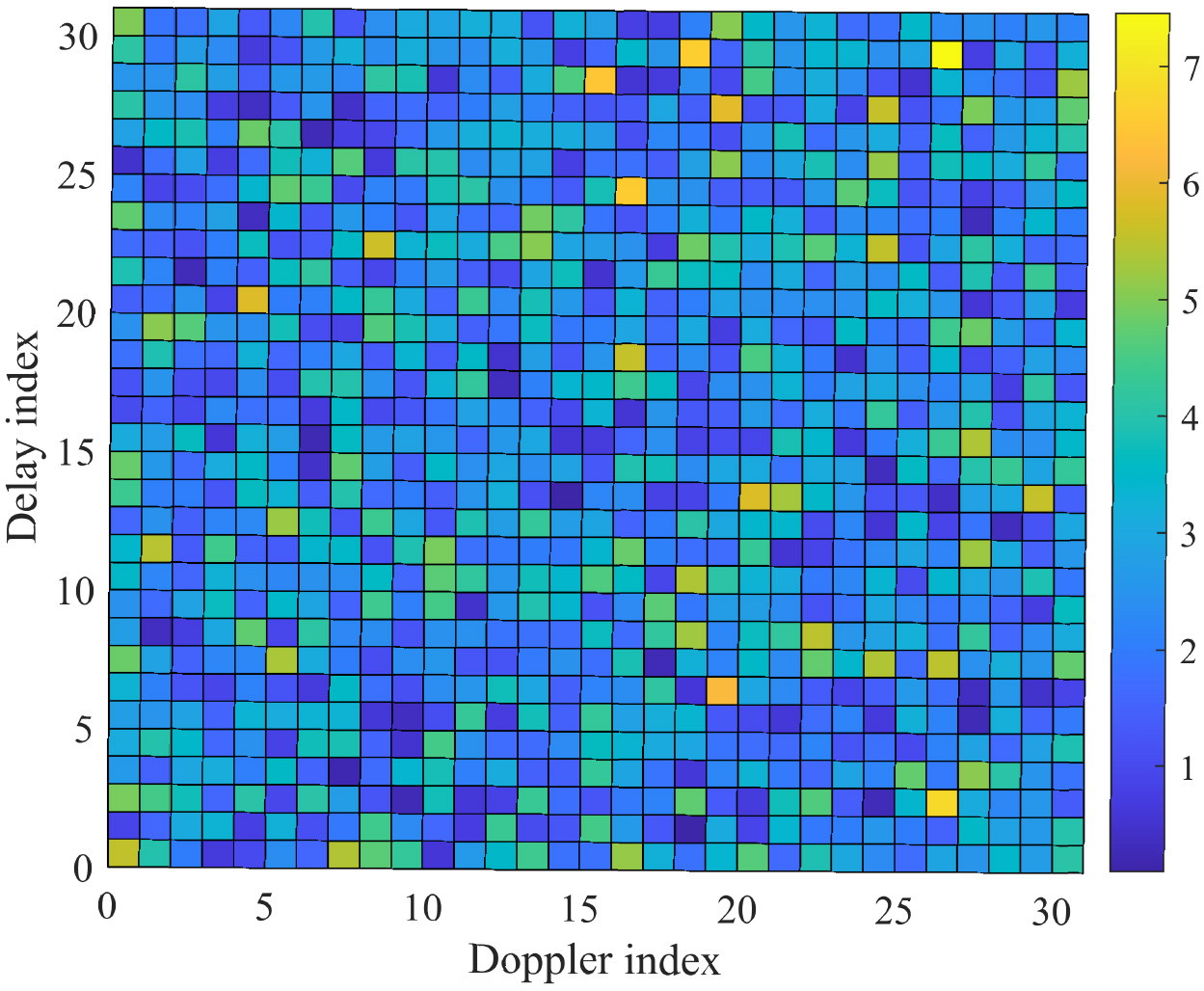}
    }
    \subfigure[After performing the 2D correlation.] { \label{correlation_fig}
        \includegraphics[width=0.46\columnwidth]{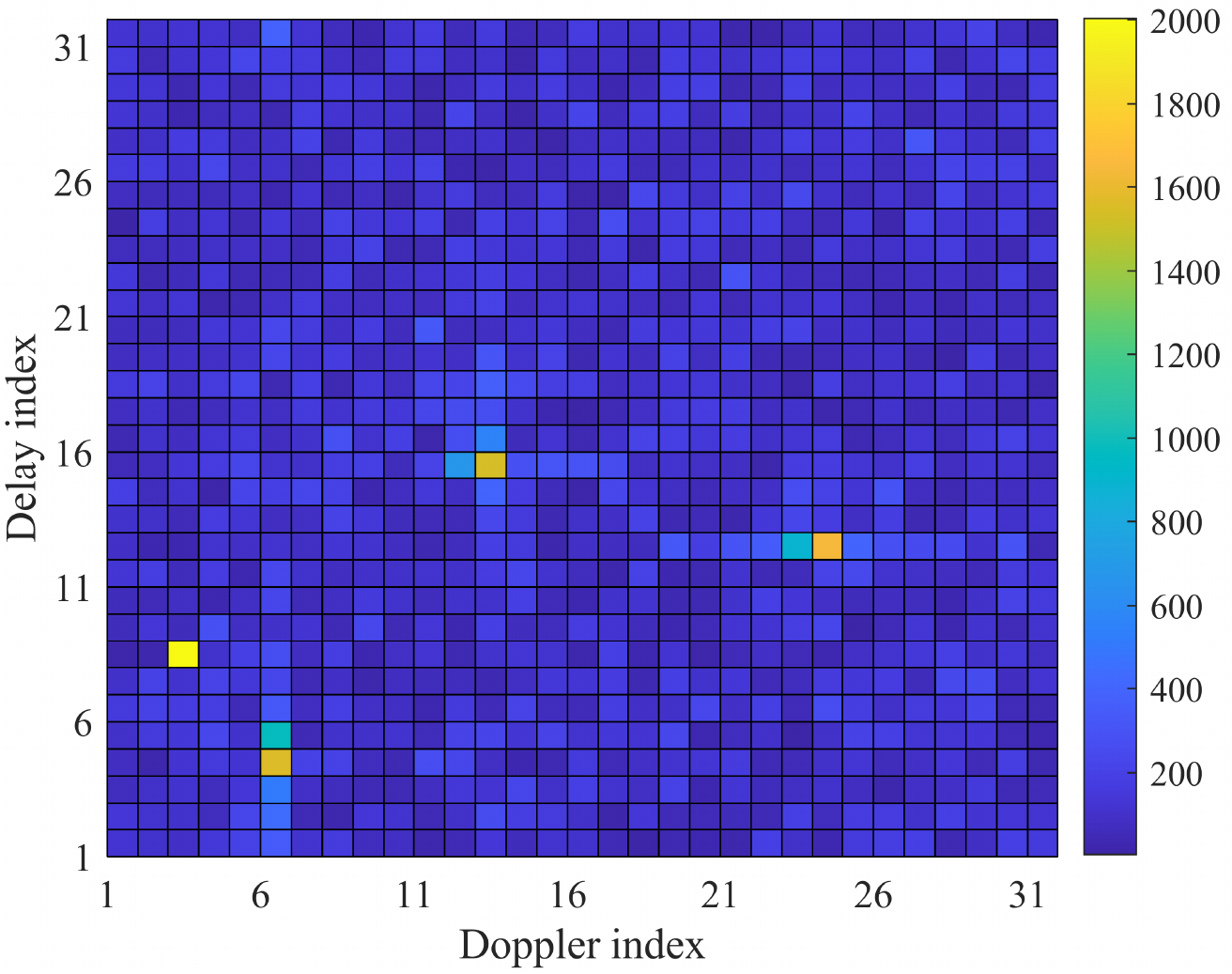}
    }
    \caption{The delay-Doppler matrices for radar sensing via OTFS waveform before and after performing the 2D correlation, where $P=4$ targets are considered and the delay and Doppler indices are of fractional values}
    \label{matrix_figure}
\end{figure}
\subsection{2D Correlation-based Operation}
We will describe the 2D correlation-based parameter estimation algorithm in this subsection. Though the received OTFS symbol matrix is represented in the delay and Doppler domain, we cannot localize the targets-of-interest directly from $\mathbf{Y}_{\text{DD}}$ due to the existence of information symbols. After passing the time-varying channel, the DD domain symbols under different delay and Doppler bins will overlap with each other, which makes the received DD domain signals contain both the channel responses and the overlapped responses from the DD domain information symbols. Inspired by pulse compression radar sensing, a 2D correlation-based estimator, which can be considered as pulse compression along both the delay and Doppler axes, is implemented to obtain the delay and Doppler parameters of the targets.
\vspace{-0.1em}
Denote the matrix after 2D pulse compression (correlation) as $\mathbf{V}$, then the accumulated correlation coefficient under different delay and Doppler indices can be expressed as
\begin{equation}
    \label{correlation_DD}
    \begin{aligned}
        V[k,l]=\sum_{n=0}^{N-1} & \sum_{m=0}^{M-1}Y_{\text{DD}}^{*}[n,m]               \\
                                & \times X_{\text{DD}}\left[[n-k]_{N},[m-l]_{M}\right]
    \end{aligned}
\end{equation}
\vspace{-0.1em}
To have a better illustration of the proposed method, we establish a toy example of OTFS radar sensing. In this example, we assume there are $P=4$ targets, and the normalized quadrature phase shift keying (QPSK) information symbols are generated randomly. Let set both $M$ and $N$ to be 32, and the delay and Doppler indices associated with different targets are set as $\{14.29, 7, 3.37, 11.12\}$ and $\{11.72, 2, 5.06, 22.65\}$, respectively. The DD domain received symbol matrix under this specific scenario is shown in Fig.\ref{received_matrix}, and the matrix after pulse compression is represented in Fig.\ref{correlation_fig}. From Fig.\ref{received_matrix}, the received symbol matrix $Y_{\text{DD}}$ is dense due to the overlapped responses from DD domain information symbols. It is hard to identify the targets from this matrix. However, after performing a 2D correlation operation, the targets are more localized. We can view this procedure as a special DD domain pulse compression, which has a similar function in radar sensing to enhance the acquisition of delay and Doppler responses.
\subsection{Target Detection via Generalized Likelihood Ratio Test}\label{target_detection}
\vspace{-0.1em}
After obtaining the 2D correlation matrix, the number of targets, i.e., $P$, remains unknown. Hence, we propose an algorithm to detect the targets sequentially. Once a target is detected, the integer parts of the corresponding parameters can be estimated at the same time. When no more targets are detected, the algorithm stops.

Before introducing the details of the target detection algorithm, let's first rewrite (\ref{correlation_DD}) in an alternative matrix form, which will make the representation of the detection procedure clearer. By substituting (\ref{effective_channel}) and (\ref{output_DD}) into (\ref{correlation_DD}), after some straightforward linear algebra calculations, we have
\begin{equation}
    \label{vector_v}
    \mathbf{v}=\sum_{i=1}^{P}h_{i}^{*}e^{j2\pi\frac{l_{\tau_{i}}k_{\nu_{i}}}{MN}}\widetilde{\mathbf{X}} \mathbf{W}^{\text{H}}_{(\tau_{i},\nu_{i})}\mathbf{x}^{*}+\widetilde{\mathbf{X}}^{\text{T}}\mathbf{z}^{*}
\end{equation}
where the $i$-th entry of the vector $\mathbf{x}^{*}$ is the ($k_{1},l_{1}$)-th entry of the information symbol matrix $\mathbf{X}$, and $\widetilde{\mathbf{X}}$ is the rearranged matrix of transmitted symbols, i.e., the circulant shifts of $\mathbf{X}$, the ($i, j$)-th entry of the matrix is represented as
\begin{equation}
    \widetilde{X}[i,j]=x[[k_{1}-k_{2}]_{N},[l_{1}-l_{2}]_{M}]\text{.}
\end{equation}
in which $i=k_{1}M+l_{1}$ and $j=k_{2}M+l_{2}$, and $k_{(\cdot)}\in[0, N-1]$ and $l_{(\cdot)}\in[0, M-1]$ are the row and column indices of the original information matrix $\mathbf{X}$. The $(kM+l)$-th element in vector $\mathbf{h}$ is the $(k,l)$-th entry of the effective channel $\mathbf{h}_{\omega}$, and $\mathbf{W}^{\text{H}}_{(\tau_{i},\nu_{i})}$ is the matrix form of the effective channel sampling function. The ($i,j$)-th element of matrix $\mathbf{W}$, in which $i=k_1\cdot M+l_1$ and $j=k_2\cdot M+l_2$, is given by
\begin{equation}
    \widetilde{W}[i,j]=\omega(k_2-k_1-k_{\nu_{i}},l_2-l_1-l_{\tau_{i}})\text{,}
\end{equation}
where $\omega(\nu,\tau)$ is given in (\ref{sampling_function}).

In order to detect the number of targets, we model the detection procedure as a binary hypothesis testing, in which hypothesis $\mathcal{H}_{0}$ and $\mathcal{H}_{1}$ represent the absence and presence of the $i$th target, respectively. The observation under the two hypotheses can be given by
\begin{equation}
    \label{hypothesis_test_reps}
    \mathbf{v}=\left\{
    \begin{aligned}
         & \widetilde{\mathbf{X}}\mathbf{z}^{*}                                                                            & ,\text{ under }\mathcal{H}_{0} \\
         & h^{*} e^{j2\pi\frac{l_{\tau}k_{\nu}}{MN}}\widetilde{\mathbf{X}}\mathbf{W}^{\text{H}}_{(\tau,\nu)}\mathbf{x}^{*} &                                \\
         & +\widetilde{\mathbf{X}}\mathbf{z}^{*}                                                                           & ,\text{ under }\mathcal{H}_{1}
    \end{aligned}
    \right..
\end{equation}

To solve (\ref{hypothesis_test_reps}), we treat $h$, $\tau$, and $\nu$ as deterministic unknown variables and resort to the generalized likelihood ratio test (GLRT) \cite{GLRT_Olivia}, which is expressed as
\begin{equation}
    \Lambda(\mathbf{v})=\max_{h,\tau,\nu}\frac{p(\mathbf{v}|\mathcal{H}_{1};h,\tau,\nu)}{p(\mathbf{v}|\mathcal{H}_{0})}\overset{\mathcal{H}_{1}}{\underset{\mathcal{H}_{0}}{\gtrless}}\eta
\end{equation}
where $\eta$ is the threshold. By assuming $\mathbf{z}\sim\mathcal{CN}(0,\sigma^{2})$, we have $\widetilde{\mathbf{X}}^{\text{T}}\mathbf{z}^{*}\sim\mathcal{CN}(0,\frac{\sigma^2}{MN})$. Denote $\widetilde{h}=h^{*}e^{j2\pi\frac{l_{\tau_{i}}k_{\nu_{i}}}{MN}}$, the GLRT becomes
\begin{equation}
    \label{hypothesis}
    \Lambda(\mathbf{v})=\frac{e^{-\frac{1}{\sigma^2}\min_{h,\tau,\nu}\parallel\mathbf{v}-\widetilde{h}\widetilde{\mathbf{X}}\mathbf{W}^{\text{H}}_{(\tau,\nu)}\mathbf{x}^{*}\parallel^2}}{e^{-\frac{1}{\sigma^2}\parallel\mathbf{v}\parallel^2}}\overset{\mathcal{H}_{1}}{\underset{\mathcal{H}_{0}}{\gtrless}}\widetilde{\eta}\text{.}
\end{equation}

For fixed $\tau$ and $\nu$, the solution of $\widetilde{h}$ that maximizes the numerator in (\ref{hypothesis}) is
\begin{equation}
    \widetilde{h}\approx\frac{\mathbf{x}^{\text{T}}\mathbf{W}\widetilde{\mathbf{X}}^{\text{H}}\mathbf{v}}{\parallel\mathbf{x}\parallel^2}\text{,}
\end{equation}
where the approximation is due to the approximate identity property of matrix $\frac{\widetilde{\mathbf{X}}^{\text{T}}\widetilde{\mathbf{X}}^{*}}{MN}$. The approximately equal sign becomes the equal sign when $MN$ goes to infinity, and readers can refer to \cite{WORK2} for proof details. By setting the phase offset to 0 in the proof procedure of \cite{WORK2}, it comes to the approximate identity property of $\frac{\widetilde{\mathbf{X}}^{\text{T}}\widetilde{\mathbf{X}}^{*}}{MN}$ in this paper. Given $\tilde{h}$, we get the GLRT test as follows
\begin{equation}
    \label{glrt_log}
    \displaystyle\max_{\tau,\nu}\frac{\mid\mathbf{x}^{\text{T}}\mathbf{W}(\tau,\nu)\widetilde{\mathbf{X}}^{\text{H}}\mathbf{v}\mid^2}{\sigma^2\parallel\mathbf{x}\parallel^2}\overset{\mathcal{H}_{1}}{\underset{\mathcal{H}_{0}}{\gtrless}}\widetilde{\eta}\text{.}
\end{equation}
where $\widetilde{\eta}=\log(\eta)$.

Here we use the cell-averaging constant false alarm rate (CA-CFAR) method to calculate the adaptive threshold, and the details for computing $\widetilde{\eta}(\nu,\tau)$ are given in Appendix \ref{glrt_appendix}. The targets are decided at those locations where the magnitude of the peaks exceeds the threshold, and the corresponding delay and Doppler indices are considered as the parameters associated with the targets. The algorithm for target detection is summarized in Algorithm \ref{detection_algorithm}.
\begin{algorithm}
    \caption{Target detection via GLRT}
    \label{detection_algorithm}
    \begin{algorithmic}[1]
        \REQUIRE $\mathbf{Y}_{\text{DD}},\mathbf{X}_{\text{DD}},M,N$
        \ENSURE $\mathbf{k}$, $\mathbf{l}$, $P$
        \STATE Set $P=0$.
        \FOR{$k_i=1; i\leqslant N$}
        \FOR{$l_j=1; j\leqslant M$}
        \STATE Compute the $V[k_i,l_j]$ statistic via (\ref{correlation_DD}).
        \STATE Calculate the threshold $\widetilde{\eta}(\nu_{i},\tau_{j})$ via (\ref{threshold_formula}) for every grid point ($\nu_{i},\tau_{j}$)$\in\mathbf{\Gamma}$, where $\tau_i=\frac{l_i}{M\Delta f}$, and $\nu_i=\frac{k_i}{NT}$.
        \IF {$V[k_i,l_j]>\widetilde{\eta}(\nu_{i},\tau_{j})$}
        \STATE $P\leftarrow P+1$, and declare the newly detected $P$-th target with estimated parameters ($k_i,l_j$).
        \STATE Add $k_i$ and $l_j$ into $\mathbf{k}$ and $\mathbf{l}$, respectively.
        \ENDIF
        \ENDFOR
        \ENDFOR
    \end{algorithmic}
\end{algorithm}
\subsection{Fractional Parameter Estimation}
By taking the peaks on the delay-Doppler plane $\bm{\Gamma}$, we can only get integer parameters, the fractional parts of the delay and Doppler indices remain unknown, which will lead to an inaccurate sensing result \cite{grid_off}. We will introduce a simple method to estimate the fractional part of delay and Doppler indices in this subsection.

By observing the 2D correlation matrix shown in Fig.\ref{correlation_fig}, we can see that for the targets that have a fractional delay and Doppler indices, there exists power leakage from the corresponding delay-Doppler bins to their neighbors. The explanation of the power leakage caused by the fractional indices can be found in \cite{OTFS_Window_Design}. Inspired by this obsession, we propose a difference-based method to estimate the fractional parts of indices, which is explained as follows.

Note that the delay and Doppler indices of the $i$-th target can be represented as $k_{\nu_{i}}=k_{i}+\kappa_{\nu_{i}}$ and $l_{\tau_{i}}=l_{i}+\iota_{\tau_{i}}$, respectively. Let us first consider calculating the fractional part of the Doppler index under noiseless conditions. In this paper, we assume that there are no two targets that have the same delay or Doppler shifts. After 2D correlation, the row indices of the maximum and the second maximum magnitudes at the $l_{i}$th column in the delay-Doppler matrix, i.e., $k_{\nu_{1}}^{\prime}$ and $k_{\nu_{2}}^{\prime}$ are
\begin{equation}\label{get_two_max}
    \begin{aligned}
        k_{\nu_{1}}^{\prime} & =\argmax_{k\in\left\{\left\lceil -N/2 \right\rceil,\dots,\left\lceil N/2 \right\rceil-1\right\}}|V[k,l_{i}]|\text{,}                                   \\
        k_{\nu_{2}}^{\prime} & =\argmax_{k\in\left\{\left\lceil -N/2 \right\rceil,\dots,\left\lceil N/2 \right\rceil-1\right\}\backslash\{k_{\nu_{1}}^{\prime}\}}|V[k,l_{i}]|\text{.}
    \end{aligned}
\end{equation}
Having $k_{\nu_{1}}^{\prime}$ and $k_{\nu_{2}}^{\prime}$ in hand, we have the following proposition.
\begin{proposition}
    In noiseless scenarios, the actual Doppler index $k_{i}+\kappa_{\nu_{i}}$ of the $i$-th target must fall into the interval bounded by $k_{\nu_{1}}^{\prime}$ and $k_{\nu_{2}}^{\prime}$, where $k_{i}=k_{\nu_{1}}^{\prime}$, and $|k_{\nu_{2}}^{\prime}-k_{\nu_{1}}^{\prime}|=1$. The ratio between the magnitudes of the correlation coefficients with Doppler indices $k_{\nu_{1}}^{\prime}$ and $k_{\nu_{2}}^{\prime}$ and the same delay index can be approximated by
    \begin{equation}\label{derivation}
        \frac{|V[k_{\nu_{1}}^{\prime},l_{i}]|}{|V[k_{\nu_{2}}^{\prime},l_{i}]|}\approx\frac{|k_{\nu_{2}}^{\prime}-k_{\nu_{1}}^{\prime}-\kappa_{\nu_{i}}|}{|-\kappa_{\nu_{i}}|}\text{,}
    \end{equation}
    where the approximation error is at the order of $\mathcal{O}\left(\frac{1}{MN}\right)$.
\end{proposition}
\begin{proof}
    See Appendix \ref{proof1}.
\end{proof}
Therefore, the fractional Doppler can be derived as
\begin{equation}\label{get_kappa}
    \kappa_{\nu_i}=\frac{(k_{\nu_{2}}^{\prime}-k_{\nu_{1}}^{\prime})|V[k_{\nu_{2}}^{\prime},l_{i}]|}{|V[k_{\nu_{1}}^{\prime},l_{i}]|+|V[k_{\nu_{2}}^{\prime},l_{i}]|}\text{,}
\end{equation}
Similarly, by applying the above derivation to the fractional delay taps, we can get
\begin{equation}\label{get_iota}
    \iota_{\tau_i}=\frac{(l_{\tau_{2}}^{\prime}-l_{\tau_{1}}^{\prime})|V[k_{i},l_{\tau_{2}}^{\prime}]|}{|V[k_{i},l_{\tau_{1}}^{\prime}]|+|V[k_{i},l_{\tau_{2}}^{\prime}]|}\text{,}
\end{equation}
where $l_{\tau_{1}}^{\prime}$ and $l_{\tau_{2}}^{\prime}$ are the column indices of the maximum and the second maximum magnitudes in the $k_{i}$th row of the 2D correlation matrix $\mathbf{V}$.

The algorithm to estimate the fractional parts of the delay and Doppler indices are summarized in Algorithm, where $\boldsymbol{k}=[k_{1},\dots,k_{P}]$ and $\boldsymbol{l}=[l_{1},\dots,l_{P}]$ denotes the integer parts of the Doppler and delay indices estimated from the target detection algorithm described in Section \ref{target_detection}, respectively.
\begin{algorithm}
    \caption{Estimate the fractional delay and Doppler indices from 2D correlated delay-Doppler matrix}
    \label{estimate_algorithm}
    \begin{algorithmic}[1]
        \REQUIRE $\mathbf{Y}_{\text{DD}},\mathbf{X}_{\text{DD}},M,N,P$
        \ENSURE $\boldsymbol{\kappa_{\nu}}$, $\boldsymbol{\iota_{\tau}}$
        \STATE Compute $\mathbf{V}$ via (\ref{correlation_DD}).
        \STATE Pick the largest $P$ peaks in $\mathbf{V}$, get $\boldsymbol{k}$ and $\boldsymbol{l}$
        \FOR{$i=1; i\leqslant P$}
        \STATE $k_{i}\gets k_{\nu_{1}}$
        \IF {$|V[[k_{i}+1]_{N}, l_{i}]|>|V[[k_{i}-1]_{N}, l_{i}]|$}
        \STATE $k_{\nu_{2}}\gets[k_{i}+1]_{N}$
        \ELSE
        \STATE $k_{\nu_{2}}\gets[k_{i}-1]_{N}$
        \ENDIF
        \STATE Calculate $\kappa_{\nu_{i}}$ using (\ref{get_kappa})
        \STATE $l_{i}\gets l_{\tau_{1}}$
        \IF {$|V[k_{i},[l_{i}+1]_{N}]|>|V[k_{i}, [l_{i}-1]_{M}]|$}
        \STATE $l_{\tau_{2}}\gets[l_{i}+1]_{M}$
        \ELSE
        \STATE $l_{\tau_{2}}\gets[l_{i}-1]_{M}$
        \ENDIF
        \STATE Calculate $\iota_{\tau_{i}}$ using (\ref{get_iota})
        \ENDFOR
    \end{algorithmic}
\end{algorithm}
\subsection{Cramer-Rao Lower Bound (CRLB)}
In this subsection, we derive the CRLB of the delay and Doppler estimation to evaluate the proposed algorithm. By substituting (\ref{sampling_function}) and (\ref{effective_channel}) into (\ref{output_DD}), we can rewrite (\ref{output_DD}) as
\begin{equation}
    Y_{\text{DD}}[k,l]=U_{\text{DD}}[k,l]+Z_{\text{DD}}[k,l]\text{,}
\end{equation}
where $\mathbf{U}_{\text{DD}}$ is the matrix of transmitted DD domain symbols under a noiseless scenario, whose ($k,l$)-th entry is
\begin{align}
     & U_{\text{DD}}[k,l]=\sum_{i=1}^{P}\sum_{n=0}^{N-1}\sum_{m=0}^{M-1}h_{i}e^{-j2\pi\frac{(k_{i}+\kappa_{i})(l_{i}+\iota_{i})}{MN}}X_{\text{DD}}[n,m]\nonumber                     \\
     & \frac{1}{NM}\sum_{k^{\prime}=0}^{N-1}e^{-j2\pi\frac{(k-n-k_{i}-\kappa_{i})k^{\prime}}{N}}\sum_{l^{\prime}=0}^{M-1}e^{j2\pi\frac{(l-m-l_{i}-\iota_{i})l^{\prime}}{M}}\nonumber
\end{align}
We define $\bm{\theta}=[\kappa_{1},...,\kappa_{P},\iota_{1},...,\iota_{P}]^{\text{T}}$, i.e., the fractional parts of delay and Doppler indices. Then the CRLB of the $j$-th element of $\mathbf{\theta}$ is the $j$-th diagonal element of the inverse of the Fisher information matrix \cite{Zacharia2022}
\begin{equation}
    \theta^{\text{CRLB}}_{j}=\left[\textbf{\textit{I}}^{-1}(\bm{\theta})\right]_{jj}
\end{equation}
where the the ($i,j$)-th element of the Fisher information matrix $\mathbf{\textit{I}}(\bm{\theta})\in\mathbb{C}^{2P\times2P}$ is
\begin{equation}
    \label{fisher_ij}
    \left[\mathbf{\textit{I}}(\bm{\theta})\right]_{ij}=-\mathbb{E}\left[\frac{\partial^2\log f(\mathbf{y}|\bm{\theta})}{\partial\theta_{i}\partial\theta_{j}}\right]
\end{equation}
where $i=1,...,2P$ and $j=1,...,2P$, and the expectation is taken with respect to $f(\mathbf{y}|\bm{\theta})$, which is given by
\begin{equation}
    f(\mathbf{y}|\bm{\theta})=\prod_{i^{\prime}=1}^{MN}\frac{1}{\sqrt{2\pi\sigma^2}}e^{-\frac{|y_{i^{\prime}}-u_{i^{\prime}}|^2}{2\sigma^2}}
\end{equation}
where $y_{i^{\prime}}$ and $u_{i^{\prime}}$ are the $i^{\prime}$-th entry of the vectors $\mathbf{y}$ and $\mathbf{u}$, in which $i^{\prime}=kM+l$, i.e., the ($k,l$)-th entries of the two matrixes $\mathbf{Y}_{\text{DD}}$ and $\mathbf{U}_{\text{DD}}$. Then the logarithm of the likelihood function $\log f(\mathbf{y}|\bm{\theta})$ is
\begin{equation}
    \label{log_func}
    \log f(\mathbf{y}|\bm{\theta})=-\frac{MN}{2}\log(2\pi\sigma^2)-\frac{1}{2\sigma^2}\sum_{i^{\prime}=1}^{MN}|y_{i^{\prime}}-u_{i^{\prime}}|^2
\end{equation}
Substitute (\ref{log_func}) into (\ref{fisher_ij}), the ($i,j$)-th entry of the Fisher matrix becomes
\begin{equation}
    \left[\mathbf{\textit{I}}(\bm{\theta})\right]_{ij}=-\frac{1}{2\sigma^2}\sum_{i^{\prime}=1}^{MN}\left[\frac{\partial u_{i^{\prime}}}{\partial\theta_{i}}\frac{\partial u_{i^{\prime}}^{*}}{\partial\theta_{j}}+\frac{\partial u_{i^{\prime}}^{*}}{\partial\theta_{i}}\frac{\partial u_{i^{\prime}}}{\partial\theta_{j}}\right]
\end{equation}

The first derivation of $u_{i^{\prime}}$ towards the $p$-th projection of $\bm{\theta}$, $\frac{\partial u_{i^{\prime}}}{\partial\theta_{p}}$, is expressed in (\ref{CRLB}), in which $p^{\prime}$ is the result of $p$ modulo $P$. The CRLB is normalized in order to evaluate the performance of the proposed algorithm in terms of the normalized mean square error (NMSE). The CRLB for the fractional delay and Doppler indices are defined by
\begin{align}
     & \kappa_{\text{bound}}=\frac{\sum_{j=1}^{P}\theta_{j}^{\text{CRLB}}}{\parallel\bm{\kappa}_{\bm{\theta}}\parallel_{2}^{2}},  \\
     & \iota_{\text{bound}}=\frac{\sum_{j=P+1}^{2P}\theta_{j}^{\text{CRLB}}}{\parallel\bm{\iota}_{\bm{\theta}}\parallel_{2}^{2}},
\end{align}
where $\bm{\kappa}_{\bm{\theta}}=[\kappa_{1},...,\kappa_{P}]^{\text{T}}$ and $\bm{\iota}_{\bm{\theta}}=[\iota_{1},...,\iota_{P}]^{\text{T}}$.
\begin{table*}[h]
    \small\begin{equation}
        \label{CRLB}
        \frac{\partial u_{i^{\prime}}}{\partial\theta_{p}}=\left\{
        \begin{aligned}
             & \sum_{n=0}^{N-1}\sum_{m=0}^{M-1}h_{p^{\prime}}X_{\text{DD}}[n,m]\left(\frac{-j2\pi(l_{p^{\prime}}+\iota_{p^{\prime}})}{MN}e^{-j2\pi\frac{(l_{p^{\prime}}+\iota_{p^{\prime}})(k_{p^{\prime}}+\kappa_{p^{\prime}})}{MN}}\omega(k-n-k_{p^{\prime}}-\kappa_{p^{\prime}},l-m-l_{p^{\prime}}-\iota_{p^{\prime}})\right.                                          \\
             & \left.+e^{-j2\pi\frac{(l_{p^{\prime}}+\iota_{p^{\prime}})(k_{p^{\prime}}+\kappa_{p^{\prime}})}{MN}}\sum_{k^{\prime}=0}^{N-1}\frac{j2\pi k^{\prime}}{N}e^{-j2\pi\frac{(k-n-k_{p^{\prime}}-\kappa_{p^{\prime}})k^{\prime}}{N}}\sum_{l^{\prime}=0}^{M-1}e^{j2\pi\frac{(l-m-l_{p^{\prime}}-\iota_{p^{\prime}})}{M}}\right)\text{, }1\leq p\leq P               \\
             & \sum_{n=0}^{N-1}\sum_{m=0}^{M-1}h_{p^{\prime}}X_{\text{DD}}[n,m]\left(\frac{-j2\pi(k_{p^{\prime}}+\kappa_{p^{\prime}})}{MN}e^{-j2\pi\frac{(l_{p^{\prime}}+\iota_{p^{\prime}})(k_{p^{\prime}}+\kappa_{p^{\prime}})}{MN}}\omega(k-n-k_{p^{\prime}}-\kappa_{p^{\prime}},l-m-l_{p^{\prime}}-\iota_{p^{\prime}})\right.                                         \\
             & \left.+e^{-j2\pi\frac{(l_{p^{\prime}}+\iota_{p^{\prime}})(k_{p^{\prime}}+\kappa_{p^{\prime}})}{MN}}\sum_{l^{\prime}=0}^{M-1}\frac{-j2\pi l^{\prime}}{M}e^{j2\pi\frac{(l-m-l_{p^{\prime}}-\iota_{p^{\prime}})l^{\prime}}{M}}\sum_{k^{\prime}=0}^{N-1}e^{-j2\pi\frac{(k-n-k_{p^{\prime}}-\kappa_{p^{\prime}})k^{\prime}}{N}}\right)\text{, }P+1\leq p\leq 2P
        \end{aligned}
        \right.
    \end{equation}\normalsize
\end{table*}
\section{Simulation Results}
\begin{figure} \centering
    \subfigure[The RMSE of range estimation.] {\label{range_compare}
        \includegraphics[width=0.45\columnwidth]{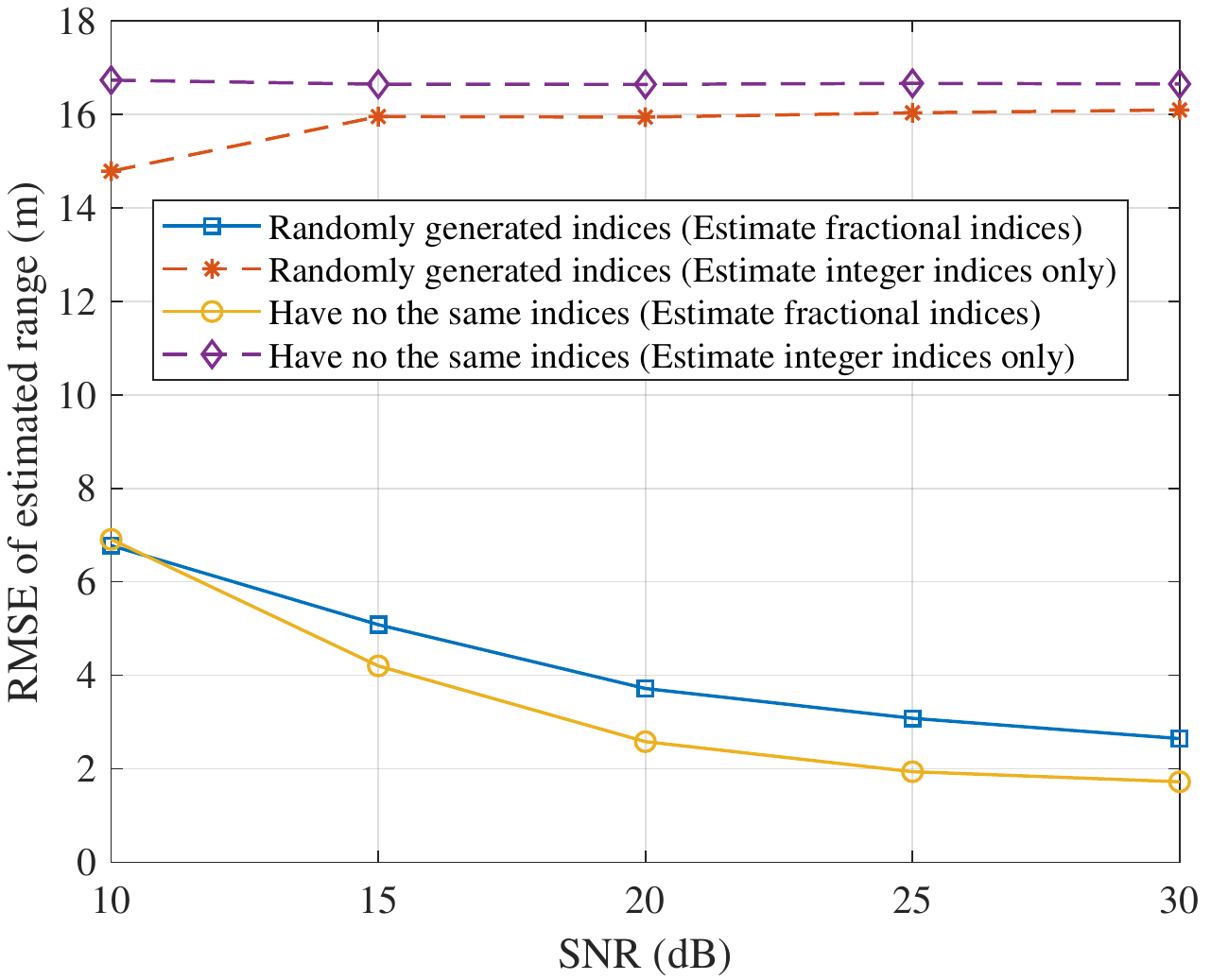}
    }
    \subfigure[The RMSE of velocity estimation.] { \label{velocity_compare}
        \includegraphics[width=0.45\columnwidth]{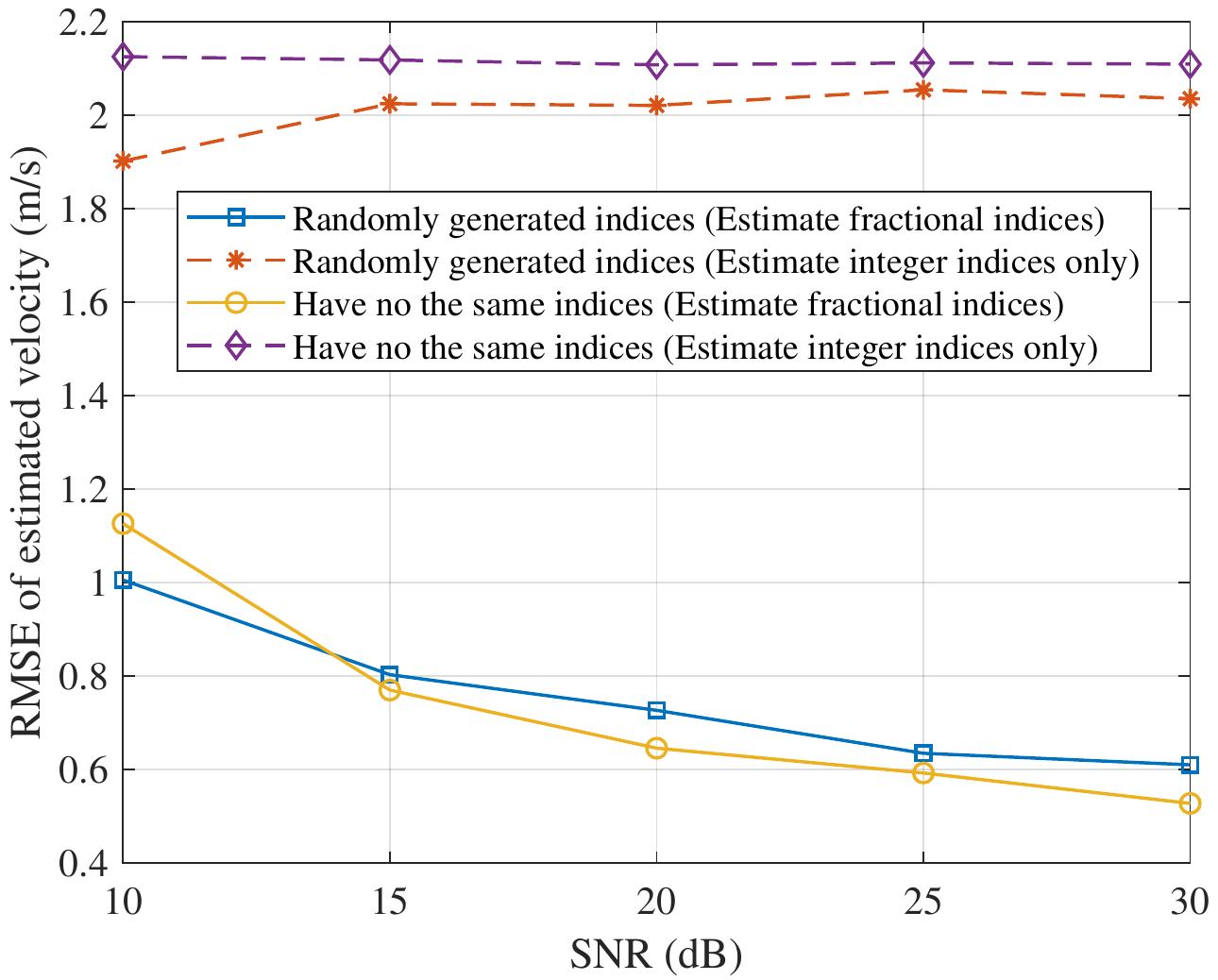}
    }
    \caption{The comparision between the estimated RMSE of target parameters under the scenario where there are no targets have the same dealy or Doppler indices and the scenario where the parameters of targets are generated randomly, where $P=4$ targets are considered.}
    \label{compare_figure}
\end{figure}

In this section, we investigate the estimation performance under various conditions through Monter Carlo simulations. The simulation results are averaged from $10^4$ OTFS frames, and each OTFS frame has $N=64$ time slots and $M=128$ subcarriers in the TF domain. The information bits carried by the OTFS frame are generated randomly and mapped to QPSK symbols. The carrier frequency is set as 24 GHz with 39.063 kHz subcarrier spacing. The maximum speed of the target is set to be 440 km/h with a maximum range of 3830 m. We consider two sensing scenarios in the simulation, in which there are $P=4$ and $P=6$ targets, respectively. We compare the performance of our proposed estimation algorithm with the conventional periodogram-based estimation algorithm via OFDM sensing, in which the number of subcarriers and time slots for OFDM sensing is set the same as the OTFS sensing to guarantee the same calculation complexity.
\begin{figure}\centering
    \includegraphics[width=0.9\columnwidth]{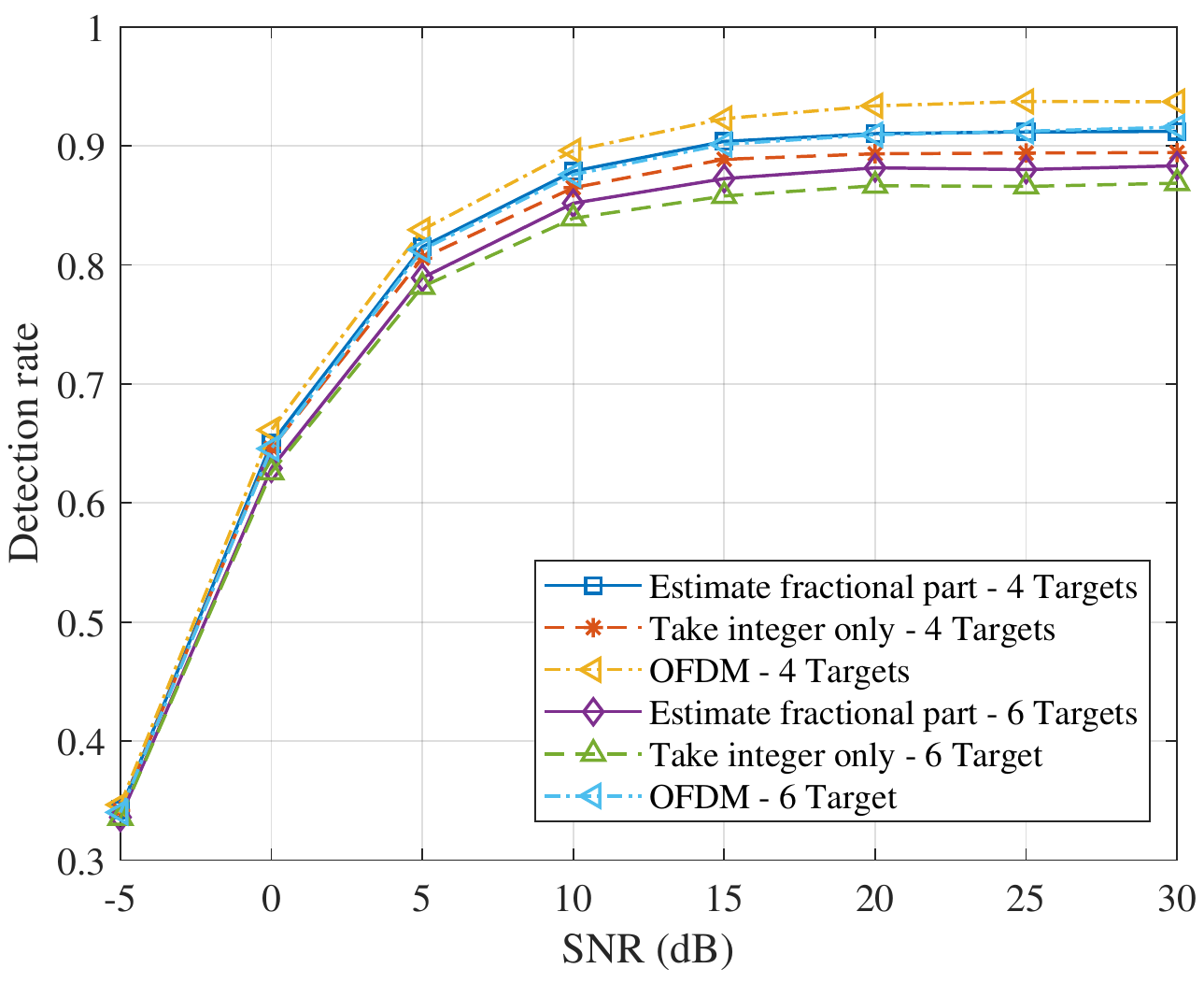}
    \caption{The detection ratio with respect to SNR.}
    \label{detection_rate}
\end{figure}

\begin{figure}\centering
    \includegraphics[width=0.9\columnwidth]{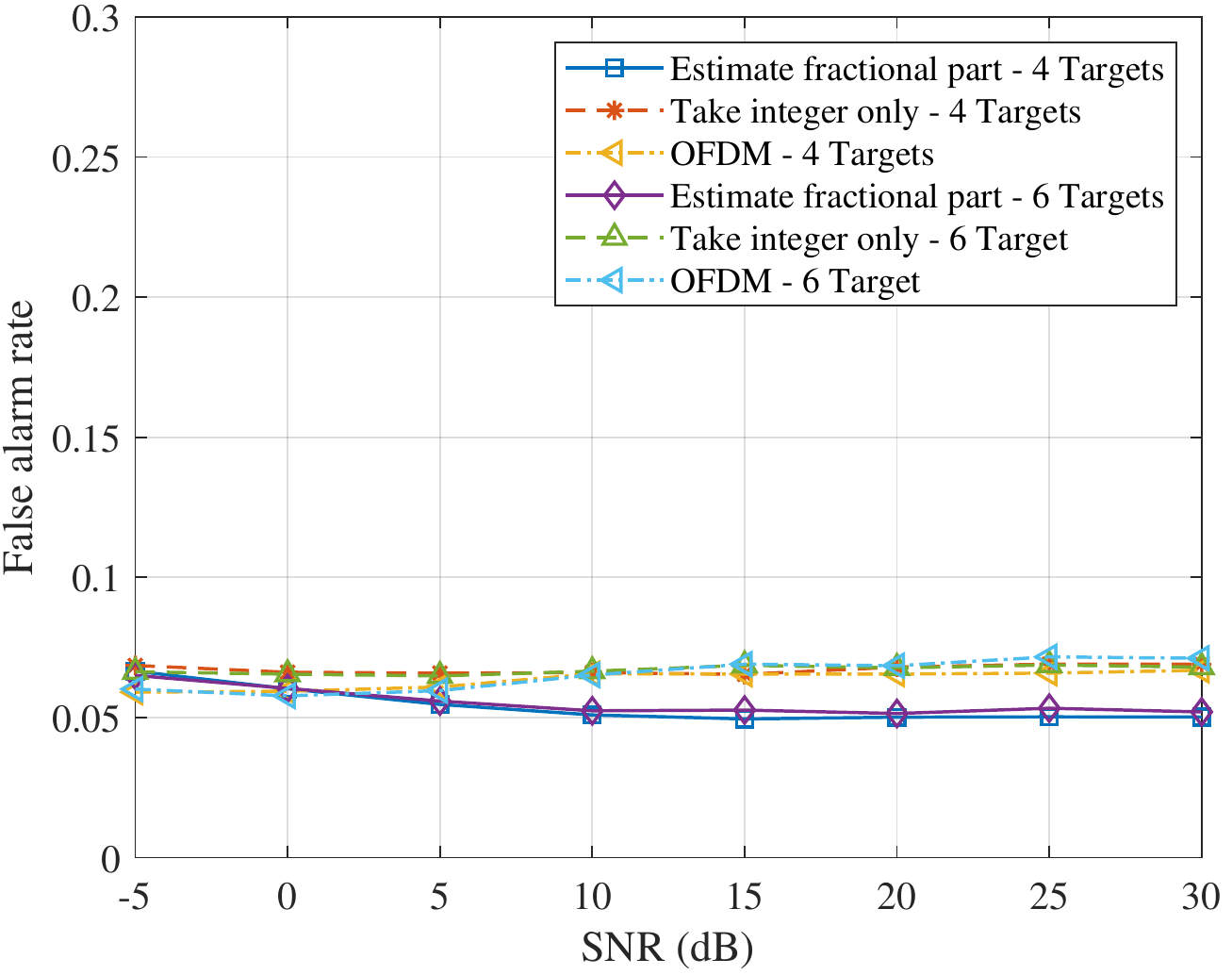}
    \caption{The false alarm rate with respect to SNR.}
    \label{far}
\end{figure}

The detection rate and the false alarm rate are counted and shown in Figure \ref{detection_rate} and Figure \ref{far}, respectively. The root-mean-square errors (RMSEs) of estimated range and velocity are calculated and drawn versus signal-to-noise ratio (SNR) in Figure \ref{range_rmse} and Figure \ref{velocity_rmse}, respectively. The comparisons between the NMSE and the theoretical CRLB of the estimations for range and velocity are shown in Figure \ref{range_crlb} and Figure \ref{velocity_crlb}.

Since the proposed parameter estimation algorithm is proved under the assumption that no two targets have the same Doppler or delay, we want to show the performance of the proposed algorithms is guaranteed even though there exist targets that have the same delay or Doppler indices. From the simulation results shown in Figure \ref{compare_figure}, we can conclude that, although there is a little performance degradation, the estimation performance under our assumptions is similar to the scenario where the target parameters are generated randomly (i.e., there exist targets that have the same delay or Doppler indices). Meanwhile, we can see that the proposed fractional parameter algorithm always works better than only estimation the integer part of the delay and Doppler shifts of targets, which shows the effectiveness of the proposed algorithm.
\begin{figure}\centering
    \includegraphics[width=0.9\columnwidth]{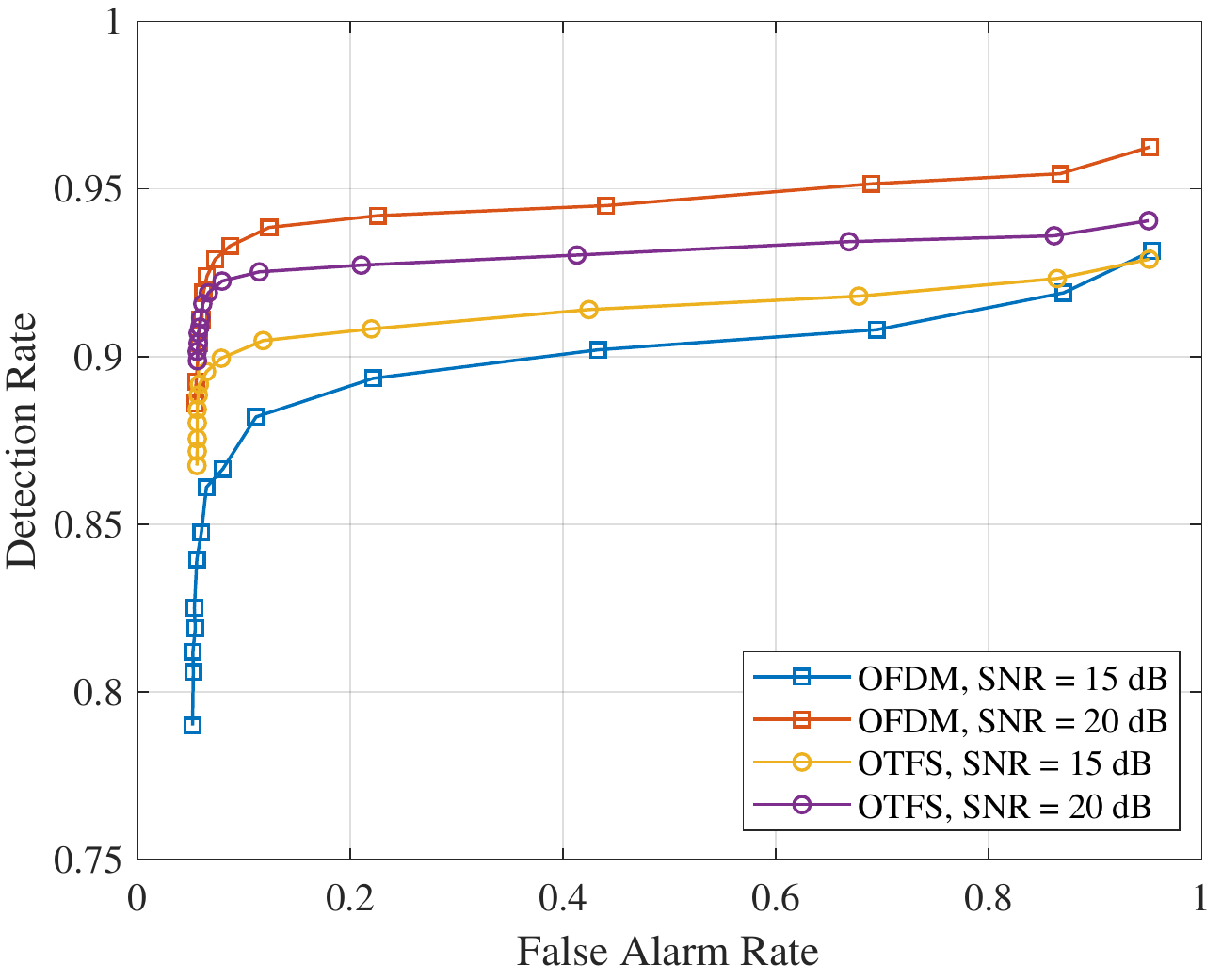}
    \caption{The ROC curve of the implemented CA-CFAR detector.}
    \label{roc_curve}
\end{figure}

It can be observed in Figure \ref{detection_rate} that with the increment of SNR, the detection rate grows. The detector designed in this paper is a false alarm rate detector, and it can be seen from Figure \ref{far} that the false alarm rate remains nearly constant under different SNRs. The receiver operating characteristic curve (ROC) is showed in Figure \ref{roc_curve} to show the effectiveness of the implemented CA-CFAR detector. The simulation results show that the performance of the implemented CA-CFAR detector is similar on both the OTFS and OFDM sensing.
\begin{figure}\centering
    \includegraphics[width=0.9\columnwidth]{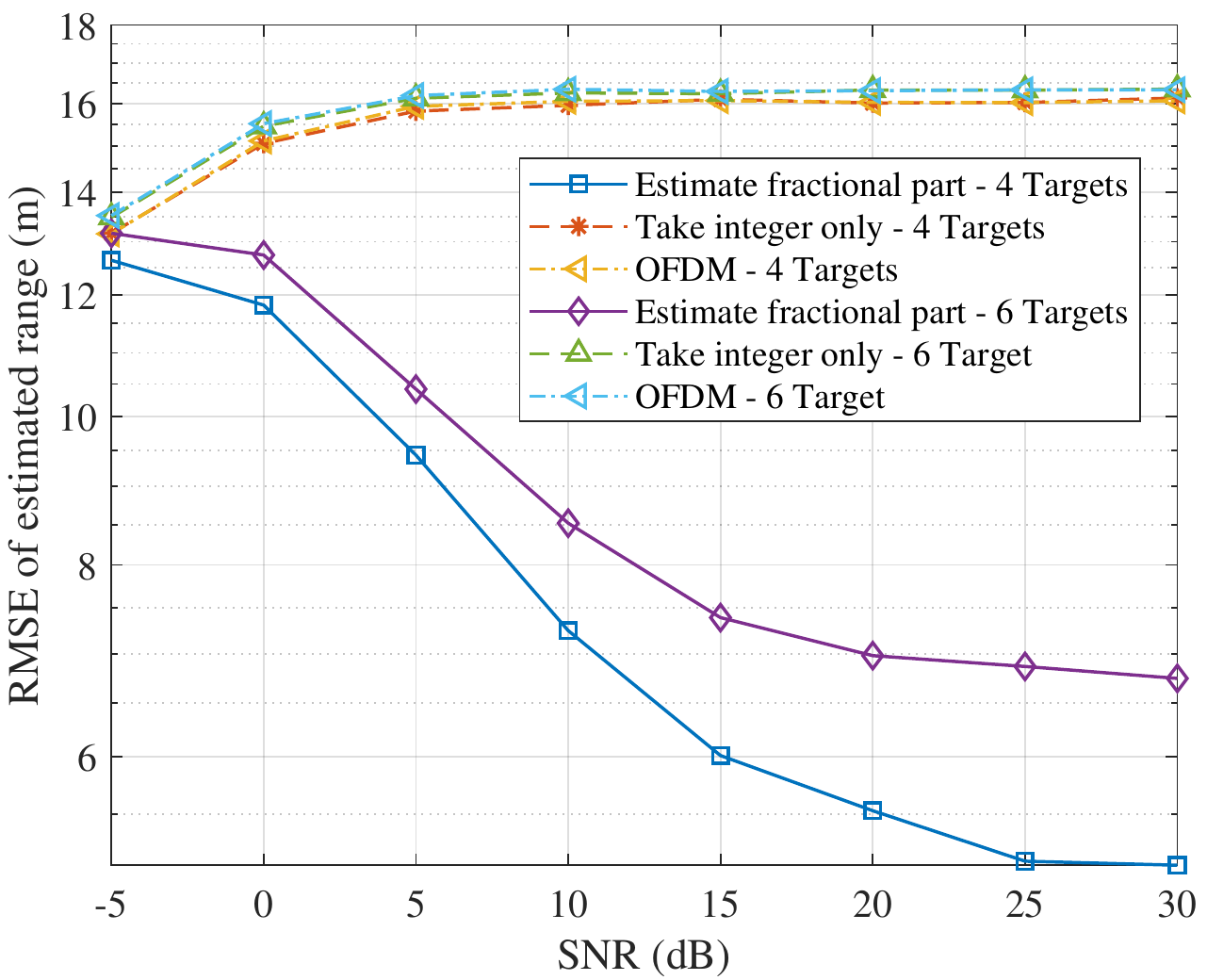}
    \caption{The RMSE of range estimation with respect to SNR.}
    \label{range_rmse}
\end{figure}

As shown in Figure \ref{range_rmse} and Figure \ref{velocity_rmse}, the RMSE increase with the increment of SNR, and it becomes a stable value when the SNR is high, which is counterintuitive. Nonetheless, this phenomenon is reasonable due to the following reasons.
\begin{figure}\centering
    \includegraphics[width=0.9\columnwidth]{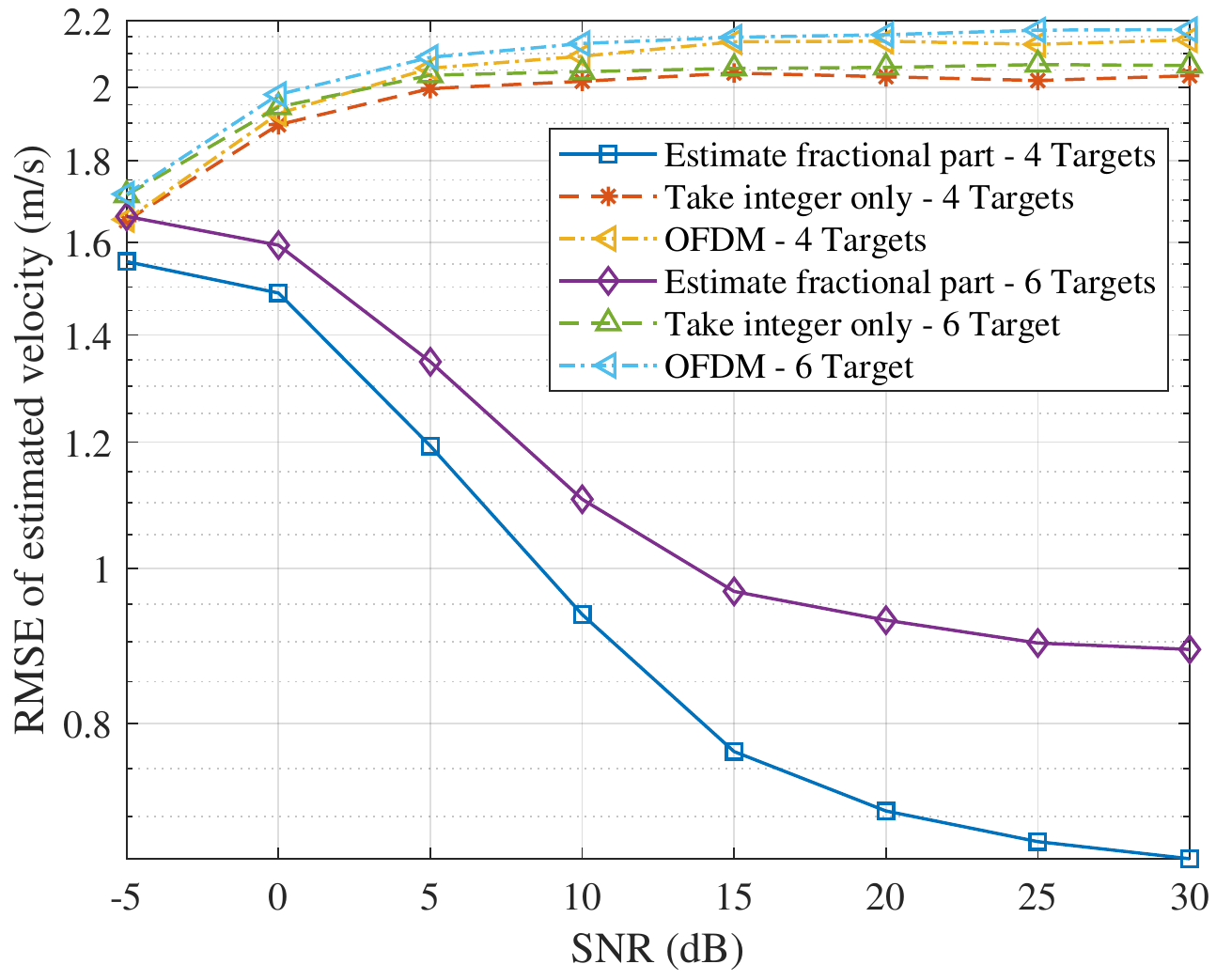}
    \caption{The RMSE of velocity estimation with respect to SNR.}
    \label{velocity_rmse}
\end{figure}

Firstly, radar systems are typically considered invalid if their detection performance is inadequate under low SNR conditions. Thus, the RMSE is not calculated in such cases. However, determining the level of detection probability at which to commence calculating the RMSE is a challenging task. Therefore, we adopt an approach to calculate the RMSE regardless of the detection probability, as long as the target is detected. Although this approach may produce counterintuitive results, it is a reasonable way to handle this problem.

Secondly, the rectangular window used in the effective channel introduces small magnitudes of entries when the fractional parts of Doppler indices are closer to 0.5. Such targets cannot be detected due to the low effective channel magnitudes and high noise levels. Only the targets whose delay and Doppler indices are close to an integer can be detected. Under these conditions, estimating only the integer parts of the parameters will yield lower RMSE. As the SNR increases, the interference from noise decreases, allowing more targets whose fractional parts of parameters are close to 0.5 to be detected. However, the RMSE remains the same due to limited resolution if only the integer part of the Doppler shift is estimated.

Therefore, the counterintuitive increase of RMSE with SNR is a result of these factors. More detailed information about the characteristics of the rectangular window can be found in Figure 5 of \cite{OTFS_Window_Design}.
\begin{figure}\centering
    \includegraphics[width=0.9\columnwidth]{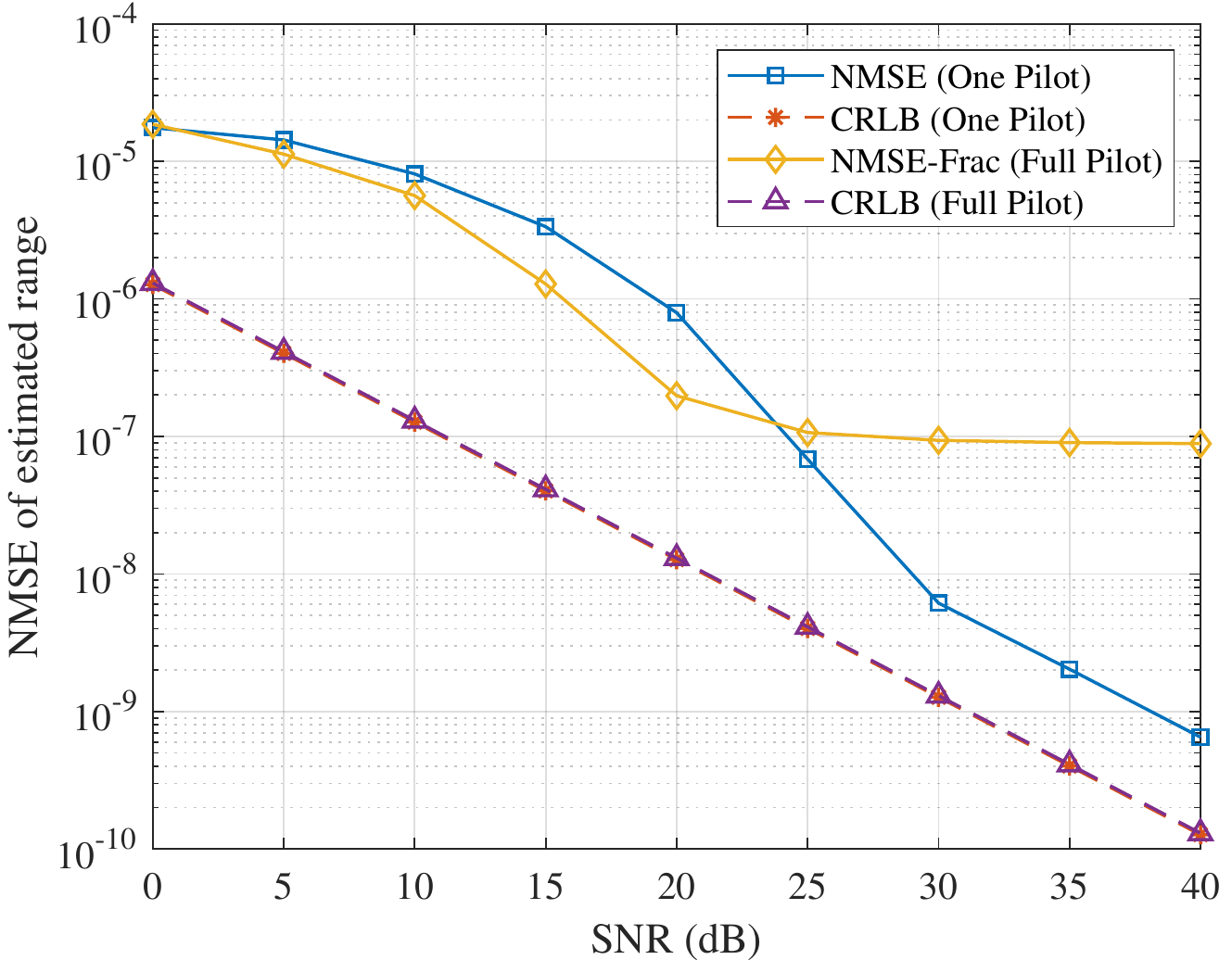}
    \caption{The NMSE and the CRLB of range estimation with respect to SNR when there are $P=4$ targets.}
    \label{range_crlb}
\end{figure}

In addition, after implementing the proposed fractional parameter estimation algorithm, the RMSE becomes lower under the same SNR compared to the simple estimation method that takes the integer parts only. The RMSE decreases since the interference caused by noise reduce with the increment of SNR.

We observe that the NMSE of the estimated parameters only decreases slightly under high SNR conditions. The sub-optimal performance in these high SNR regions is attributed to the interference caused by the overlapped responses of the data symbols after passing through the channel. This interference can be categorized as inter-symbol interference in the DD domain, as each element of the transmitting matrix $\mathbf{X}{\text{DD}}$ is assigned a modulated data symbol, resulting in none-zero values. For brevity, we refer to this element placement scheme in $\mathbf{X}{\text{DD}}$ as the 'Full Pilot' strategy (RMSE is calculated under this scenario).

In contrast, if only one DD-domain pilot is transmitted in an OTFS frame (i.e., all other elements are zero except the pilot), the NMSE approaches the Cramér-Rao lower bound (CRB). We denote this element placement scheme in $\mathbf{X}{\text{DD}}$ as the 'One Pilot' strategy for brevity. Figures 8 and 9 demonstrate that the NMSE under the 'One Pilot' strategy closely approximates the CRB in high SNR regions. This is because when only one pilot is present in $\mathbf{X}{\text{DD}}$, the inter-symbol interference in the DD domain is eliminated, allowing the NMSE to approach the CRB under high SNR conditions.

As part of our future work, we plan to investigate methods for eliminating the inter-symbol interference in the DD domain by leveraging the relationship between the signal responses in the DD domain and the target parameters. We will explore advanced algorithms, including Machine Learning methods, to address DD-domain inter-symbol interference elimination.
\begin{figure}\centering
    \includegraphics[width=0.9\columnwidth]{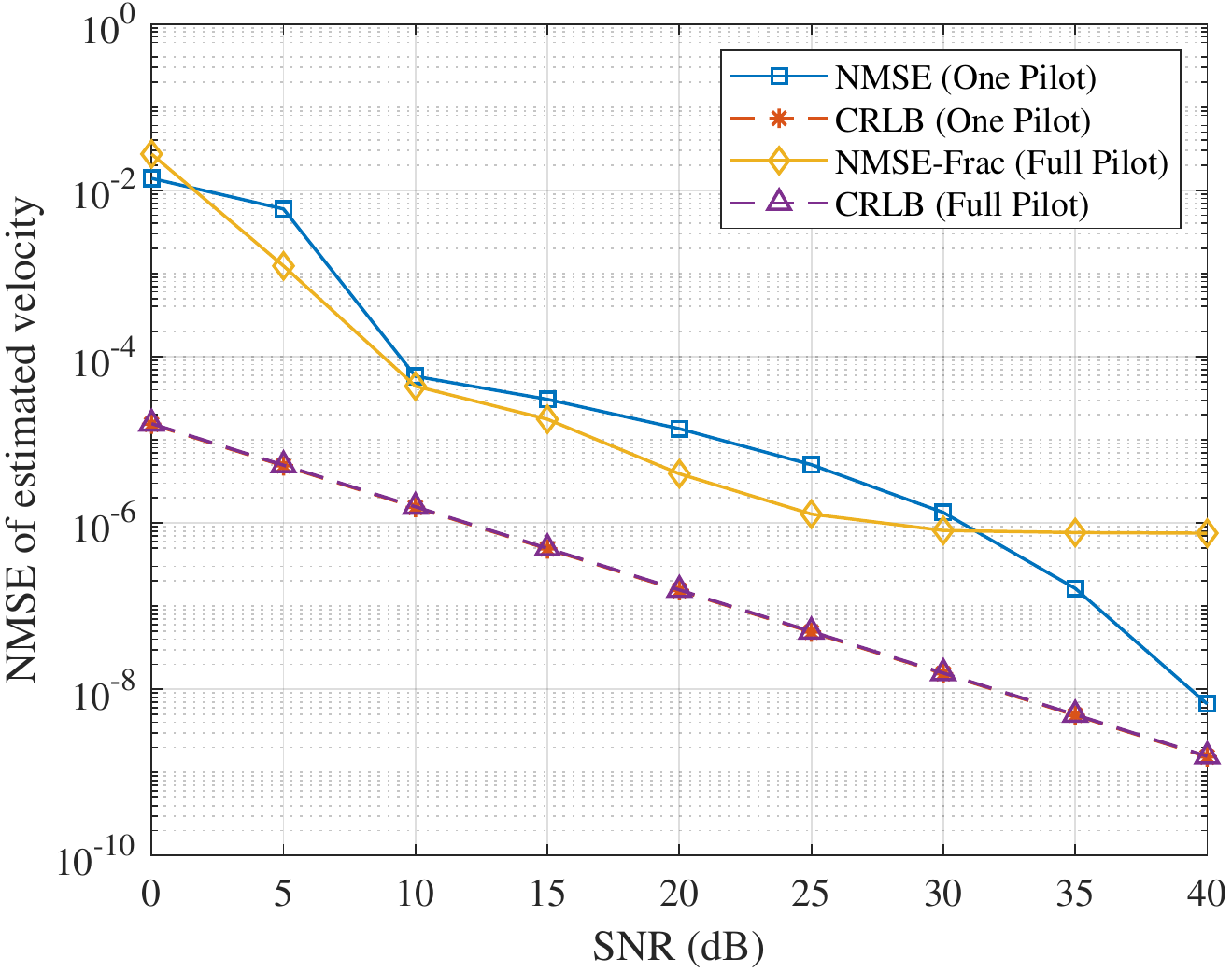}
    \caption{The NMSE and the CRLB of velocity estimation with respect to SNR when there are $P=4$ targets.}
    \label{velocity_crlb}
\end{figure}

\section{Conclusions}\label{conclusion}
In this paper, we proposed a two-step algorithm for the fractional delay and Doppler shifts estimation. Since the delay and Doppler shifts of the wireless channel can result in overlapped responses of information symbols, it is infeasible to localize the target from the received DD domain matrix directly. To obtain the parameters of targets, a 2D correlation scheme was first performed between the DD domain received and the transmitted data symbol matrix, and the integer part of delay and Doppler shifts can be obtained. After that, we implemented a difference-based method to estimate the fractional parts of the parameters. Since the number of potential targets is usually unknown in practice, we proposed a GLRT-based target detection method to get the number of targets. The simulation results show that the proposed method can detect the number of targets in the sensing scenario with a high detection probability and obtain the delay and Doppler shifts associated with multiple targets accurately.
\appendix
\section{Adaptive Threshold for Target Detection} \label{glrt_appendix}
The GLRT tests under hypotheses $\mathcal{H}_{0}$ and $\mathcal{H}_{1}$ are given by
\begin{equation}
    \begin{aligned}
         & S_{\mathcal{H}_{0}}(\nu,\tau)=\frac{\mid\mathbf{x}^{\text{T}}\mathbf{W}(\tau,\nu)\widetilde{\mathbf{X}}^{\text{H}}\widetilde{\mathbf{X}}\mathbf{z}^{*}\mid^2}{\sigma^2\parallel\mathbf{x}\parallel^2} \\
         & S_{\mathcal{H}_{1}}(\nu,\tau)=\frac{\mid\mathbf{x}^{\text{T}}\mathbf{W}(\tau,\nu)\widetilde{\mathbf{X}}^{\text{H}}\mathbf{v}\mid^2}{\sigma^2\parallel\mathbf{x}\parallel^2}
    \end{aligned}\text{.}
\end{equation}
Now we derive the adaptive threshold $\widetilde{\eta}(\nu,\tau)$ for the comparison towards the GLRT statistic $S(\nu,\tau)$ and decision for $\mathcal{H}_{0}$ or $\mathcal{H}_{1}$. Note that $S_{\mathcal{H}_{0}}(\nu,\tau)$ is a summation of a sequence of squared Gaussian complex random variables, which can be simplified as $\frac{|\widetilde{\mathbf{X}}\mathbf{z}^{*}|^2}{\sigma^2\parallel\mathbf{x}\parallel^2}$. Hence, it is exponentially distributed. The derivation of the adaptive threshold is represented below.

\begin{figure}\centering
    \includegraphics[width=\columnwidth]{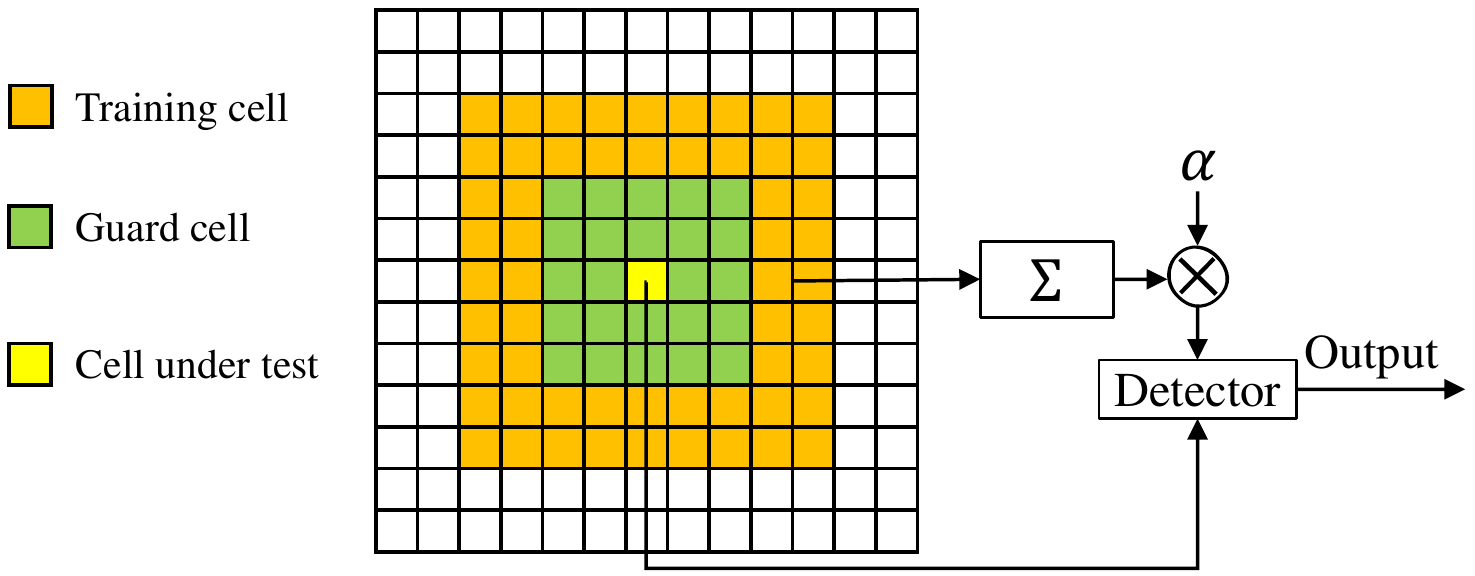}
    \caption{The illustration of 2D CA-CFAR algorithm.}
    \label{est_performance}
\end{figure}

We first derive the distribution function of $S_{\mathcal{H}_{0}}$. From (\ref{vector_v}) the value of each element in the correlated matrix can be viewed as the superposition of noise and clutters. We assume the interference of the correlated matrix entry contains two parts, i.e., the noise and the clutter, which implies the distribution of the interference follows the Rayleigh distribution. Meanwhile, since $\sigma^2\parallel\mathbf{x}\parallel^2$ is constant, it is equivalent to only compare the numerator of the GLRT statistic, which is just the value of the entry in $\mathbf{v}$ under $\mathcal{H}_{0}$, with the threshold. Denote the simplified GLRT statistic as $\widetilde{S}_{\mathcal{H}_{0}}=|\widetilde{\mathbf{X}}\mathbf{z}^{*}|^2$. Thus, the statistic $\widetilde{S}_{\mathcal{H}_{0}}$, which is a squared Rayleigh random variable, follows the exponential distribution,
\begin{equation}
    f_{\widetilde{S}_{\mathcal{H}_{0}}}(s)=\frac{1}{\widetilde{\sigma}^2}e^{-\frac{s}{\widetilde{\sigma}^2}}
\end{equation}
where $\widetilde{\sigma}^2$ is a redundant variable that can be canceled in the closed-form expression of the adaptive threshold.

Following the cell-averaging constant false alarm rate (CACFAR) procedure given in \cite{richards2014fundamentals}, we apply the 2D-CACFAR. For a fixed detection threshold $\widetilde{\eta}=\alpha\mathcal{S}_{\mathcal{H}_{0}}$, where $\alpha$ is a scaling factor needs to be calculated and $\mathcal{S}_{\mathcal{H}_{0}}$ is the mean value of training cells (shown in Fig .*), the false alarm rate is
\begin{align}
    \label{false_alarm_express}
     & P_{fa}=\mathbb{E}_{\mathcal{S}_{\mathcal{H}_{0}}}\{P[\mathcal{S}_{\mathcal{H}_{0}}>\widetilde{\eta}|\mathcal{H}_{0}]\}\nonumber                                                                                                                                       \\
     & =\mathbb{E}_{\mathcal{S}_{\mathcal{H}_{0}}}\left\{\int_{\mathcal{S}_{\mathcal{H}_{0}}}^{\infty}f_{\widetilde{S}_{\mathcal{H}_{0}}}(s)ds\right\}=E_{\mathcal{S}_{\mathcal{H}_{0}}}\{e^{-\frac{\alpha}{\widetilde{\sigma}^{2}}\mathcal{S}_{\mathcal{H}_{0}}}\}\nonumber \\
     & =M_{\mathcal{S}_{\mathcal{H}_{0}}}(u)|_{u=-\frac{\alpha}{\widetilde{\sigma}^{2}}}
\end{align}
where $M_{\mathcal{S}_{\mathcal{H}_{0}}}(u)$ is the moment generating function (MGF) of $\mathcal{S}_{\mathcal{H}_{0}}$. Denote the set of training cells as $\Xi(\nu,\tau)$, the expression of $\mathcal{S}_{\mathcal{H}_{0}}$ for a specific entry ($\nu,\tau$) on the delay-Doppler plane is
\begin{equation}
    \mathcal{S}_{\mathcal{H}_{0}}=\frac{1}{N_{s}}\left(\sum_{(\nu^{\prime},\tau^{\prime})\in\Xi(\nu,\tau)}v[\nu^{\prime},\tau^{\prime}]^2\right)
\end{equation}
where $N_{s}$ is the nubmer of training cells. It can be verified that $\mathcal{S}_{\mathcal{H}_{0}}\sim\Gamma(N_{s},\frac{\widetilde{\sigma}^2}{N_{s}})$, where $\Gamma(\alpha,\beta)$ represents for Gamma distribution. Substitute the MGF of Gamma distribution in to (\ref{false_alarm_express}), and we have
\begin{equation}
    P_{fa}=\left(\frac{1}{1+\frac{\alpha}{N_{s}}}\right)^{N_{s}}.
\end{equation}
Thus, $\alpha=N_{s}(P_{fa}^{-\frac{1}{N_{s}}}-1)$, and the threshold $\widetilde{\eta}(\nu,\tau)$ is
\begin{equation}\label{threshold_formula}
    \widetilde{\eta}(\nu,\tau)=N_{s}(P_{fa}^{-\frac{1}{N_{s}}}-1)\mathcal{S}_{\mathcal{H}_{0}}(\nu,\tau).
\end{equation}
Therefore, we declare a new target if the following condition holds
\begin{equation}
    V[k,l]>\widetilde{\eta}(\nu,\tau),
\end{equation}
where $l=\tau M\Delta f$, and $k=\nu NT$ are the delay and Doppler indices, respectively.
\section{Proof of Proposition 1}\label{proof1}
\begin{table*}
    \begin{equation}
        \label{v2_complicated}
        \resizebox{0.95\hsize}{!}{$\begin{aligned}
                     & \mathbb{E}[V[k,l]^2]=\mathbb{E}\left\{\left[\sum_{n^{\prime},m^{\prime},k^{\prime},l^{\prime}}
                    h^{*}_{\omega}[n^{\prime}-k^{\prime}, m^{\prime}-l^{\prime}]X_{\text{DD}}^{*}[k^{\prime},l^{\prime}]X_{\text{DD}}[n^{\prime}-k,m^{\prime}-l]
                    +\sum_{n^{\prime},m^{\prime}}
                    X_{\text{DD}}[n^{\prime}-k,m^{\prime}-l]Z_{\text{DD}}^{*}[n^{\prime},m^{\prime}]\right]\right.                                                                                                                                                                                                                                                                                                 \\
                     & \left.\times\left[\sum_{n^{\prime\prime},m^{\prime\prime},k^{\prime\prime},l^{\prime\prime}}
                    h^{*}_{\omega}[n^{\prime\prime}-k^{\prime\prime}, m^{\prime\prime}-l^{\prime\prime}]X_{\text{DD}}^{*}[k^{\prime\prime},l^{\prime\prime}]X_{\text{DD}}[n^{\prime\prime}-k,m^{\prime\prime}-l]
                    +\sum_{n^{\prime\prime},m^{\prime\prime}}
                    X_{\text{DD}}[n^{\prime\prime}-k,m^{\prime\prime}-l]Z_{\text{DD}}^{*}[n^{\prime\prime},m^{\prime\prime}]\right]\right\}                                                                                                                                                                                                                                                                        \\
                     & =\sum_{n^{\prime},m^{\prime},k^{\prime},l^{\prime}\atop n^{\prime\prime},m^{\prime\prime},k^{\prime\prime},l^{\prime\prime}}
                    h^{*}_{\omega}[n^{\prime}-k^{\prime}, m^{\prime}-l^{\prime}]h^{*}_{\omega}[n^{\prime\prime}-k^{\prime\prime}, m^{\prime\prime}-l^{\prime\prime}]\mathbb{E}\{X_{\text{DD}}^{*}[k^{\prime},l^{\prime}]X_{\text{DD}}[n^{\prime}-k,m^{\prime}-l]X_{\text{DD}}^{*}[k^{\prime\prime},l^{\prime\prime}]X_{\text{DD}}[n^{\prime\prime}-k,m^{\prime\prime}-l]\}                                         \\
                     & +\sum_{n^{\prime},m^{\prime},k^{\prime},l^{\prime},n^{\prime\prime},m^{\prime\prime}}h^{*}_{\omega}[n^{\prime}-k^{\prime}, m^{\prime}-l^{\prime}]\mathbb{E}\{X_{\text{DD}}^{*}[k^{\prime},l^{\prime}]X_{\text{DD}}[n^{\prime}-k,m^{\prime}-l]X_{\text{DD}}[n^{\prime\prime}-k,m^{\prime\prime}-l]Z_{\text{DD}}^{*}[n^{\prime\prime},m^{\prime\prime}]\}                                     \\
                     & +\sum_{n^{\prime\prime},m^{\prime\prime},k^{\prime\prime},l^{\prime\prime},n^{\prime},m^{\prime}}h^{*}_{\omega}[n^{\prime\prime}-k^{\prime\prime}, m^{\prime\prime}-l^{\prime\prime}]\mathbb{E}\{X_{\text{DD}}^{*}[k^{\prime\prime},l^{\prime\prime}]X_{\text{DD}}[n^{\prime}-k,m^{\prime}-l]X_{\text{DD}}[n^{\prime\prime}-k,m^{\prime\prime}-l]Z_{\text{DD}}^{*}[n^{\prime},m^{\prime}]\} \\
                     & +\sum_{n^{\prime},m^{\prime},n^{\prime\prime},m^{\prime\prime}}\mathbb{E}\{X_{\text{DD}}[n^{\prime}-k,m^{\prime}-l]X_{\text{DD}}[n^{\prime\prime}-k,m^{\prime\prime}-l]Z_{\text{DD}}^{*}[n^{\prime\prime},m^{\prime\prime}]Z_{\text{DD}}^{*}[n^{\prime},m^{\prime}]\}.
                \end{aligned}$}
    \end{equation}
    \hrule
    \vspace{-3mm}
\end{table*}

The lower and upper limits of the summation operator in what follows are omitted for brevity. In particular, the indices $n$, $n^{\prime}$, $n^{\prime\prime}$, $k$, $k^{\prime}$, and $k^{\prime\prime}$ are from $0$ to $N-1$ while $m$, $m^{\prime}$, $m^{\prime\prime}$, $l$, $l^{\prime}$, and $l^{\prime\prime}$ are from $0$ to $M-1$.

Taking the expectation $V[k,l]$ given by (\ref{correlation_DD}) yields
\begin{align}\label{E_v_kl}
     & \mathbb{E}[V[k,l]]=\sum_{n,m}\mathbb{E}\{X_{\text{DD}}[n-k,m-l]Z_{\text{DD}}^{*}[n,m]\}\nonumber                                             \\
     & + \sum_{n,m,k^{\prime},l^{\prime}}\mathbb{E}\{h^{*}_{\omega}[n-k^{\prime}, m-l^{\prime}]X_{\text{DD}}^{*}[k^{\prime},l^{\prime}] & \nonumber \\
     & \times X_{\text{DD}}[n-k,m-l]\}.
\end{align}
The expectation $\mathbb{E}[X_{\text{DD}}^{*}[k^{\prime},l^{\prime}]X_{\text{DD}}[n-k,m-l]]$ equals $0$ unless $k^{\prime}=n-k$ and $l^{\prime}=m-l$ due to the independent QPSK entries with unit power in $\mathbf{X_{\text{DD}}}$ . Meanwhile, since the information symbols are independent of the noise samples, the term $\mathbb{E}[X_{\text{DD}}[n-k,m-l]Z_{\text{DD}}^{*}[n,m]]=0$. Thus, \eqref{E_v_kl} can be simplified to be
\begin{equation}
    \mathbb{E}[V[k,l]]=MN\cdot h^{*}_{\omega}[k,l]\text{.}
\end{equation}
The variance of the entry in matrix $\mathbf{V}$ is
\begin{equation}
    \text{var}[V[k,l]]=\mathbb{E}[V[k,l]^2]-\mathbb{E}[V[k,l]]^2
\end{equation}
where $\mathbb{E}[V[k,l]^2]$ is given in (\ref{v2_complicated}). As before, only the terms with $k^{\prime}=n^{\prime}-k$, $l^{\prime}=m^{\prime}-l$, $k^{\prime\prime}=n^{\prime\prime}-k$, and $l^{\prime\prime}=m^{\prime\prime}-l$ are non zeros. Thus, the second and the third terms on the right-hand side of (\ref{v2_complicated}) can be discarded while the first and last terms are $(MN\cdot h_{\omega}^{*}[k,l])^2$ and $MN\cdot \sigma^2$, respectively.

Consequently, we have
\begin{equation}
    \mathbb{E}[V[k,l]^2]=(MN\cdot h_{\omega}^{*}[k,l])^2+MN\cdot \sigma^2\text{,}
\end{equation}
which gives the variance of $V[k,l]$, i.e., $\text{var}[V[k,l]]=MN\cdot \sigma^2$. Now, we consider the variance of $\frac{V[k,l]}{MN}$. Taking the limitation of $\text{var}[\frac{1}{MN}V[k,l]]$ gives
\begin{equation}\label{limit_v}
    \lim_{M,N\rightarrow\infty}\text{var}\left[\frac{V[k,l]}{MN}\right]=\lim_{M,N\rightarrow\infty}\left(\frac{\sigma^2}{MN}\right)=0\text{.}
\end{equation}
indicating that when $M$ and $N$ are sufficiently large, the variance of $\frac{1}{MN}V[k,l]$ vanishes. This motivates us to use $\frac{1}{MN}V[k,l]$ to approximate its expectation, i.e. $h_{\omega}^{*}[k,l]$, with an approximation error of $\mathcal{O}\left(\frac{1}{MN}\right)$.

Therefore, the ratio between the magnitudes of the correlation coefficients with the same delay index and Doppler indices $k_{\nu_{1}}^{\prime}$ and $k_{\nu_{2}}^{\prime}$ can be expressed as
\begin{align}\label{fraction_estimation}
     & \frac{|V[k_{\nu_{1}}^{\prime},l_{i}]|}{|V[k_{\nu_{2}}^{\prime},l_{i}]|}=\frac{|h_{\omega}[k_{\nu_{1}}^{\prime},l_{i}]|}{|h_{\omega}[k_{\nu_{2}}^{\prime},l_{i}]|}                                                                                                                                       \\
     & =\left\lvert \frac{\sin(-\kappa_{\nu_{i}}\pi)}{\sin(\frac{-\kappa_{\nu_i}\pi}{N})}\right\rvert \cdot \left\lvert \frac{\sin((k_{\nu_{2}}^{\prime}-k_{\nu_{1}}^{\prime}-\kappa_{\nu_{i}})\pi)}{\sin(\frac{k_{\nu_{2}}^{\prime}-k_{\nu_{1}}^{\prime}-\kappa_{\nu_{i}}}{N}\pi)}\right\rvert ^{-1}\nonumber \\
     & =\left\lvert \frac{\sin(\frac{k_{\nu_{2}}^{\prime}-k_{\nu_{v1}}^{\prime}-\kappa_{\nu_{i}}}{N}\pi)}{\sin(\frac{-\kappa_{\nu_{i}}\pi}{N})}\right\rvert \approx \frac{|k_{\nu_{2}}^{\prime}-k_{\nu_{1}}^{\prime}-\kappa_{\nu_{i}}|}{|-\kappa_{\nu_{i}}|}\nonumber\text{.}
\end{align}
Note that the small-angle approximation $\sin x \approx x$ also holds for a sufficiently large $N$. Finally, we arrive at \eqref{derivation}.
\bibliographystyle{gbt7714-numerical}
\bibliography{myref}
\end{document}